\documentclass[preprint]{aastex}

\usepackage{amssymb}
\usepackage{graphicx}
\usepackage[unicode=true]{hyperref}
\usepackage{breakurl}
\usepackage{amsmath}
\usepackage{footnote}

\shorttitle{}
\shortauthors{Hao et al.}

\begin{document}

\title{Revisiting the Relationship between the Long GRB Rate and Cosmic Star Formation History Based on a Large Swift Sample}

\author{Jing-Meng Hao\altaffilmark{1,2}, Liang Cao\altaffilmark{3,4},
You-Jun Lu\altaffilmark{3,4}, Qing-Bo Chu\altaffilmark{3,4},
Jun-Hui Fan\altaffilmark{1,2}, Ye-Fei Yuan\altaffilmark{5} and Yu-Hai Yuan\altaffilmark{1,2}}

\altaffiltext{1}{Center for Astrophysics, Guangzhou University, Guangzhou 510006, China; jmhao@gzhu.edu.cn}
\altaffiltext{2}{Astronomy Science and Technology Research Laboratory of Department of Education of Guangdong Province, Guangzhou 510006, China}
\altaffiltext{3}{National Astronomical Observatories, Chinese Academy of Sciences, 20A Datun Road, Beijing 100101, China}
\altaffiltext{4}{School of Astronomy and Space Sciences, University of Chinese Academy of Sciences, 19A Yuquan Road, Beijing 100049, China}
\altaffiltext{5}{Key Laboratory for Research in Galaxies and Cosmology CAS, Department of Astronomy, University of Science and Technology of China, Hefei, Anhui 230026, China}

\begin{abstract}
The exact relationship between the long gamma-ray burst (LGRB) rate
and the cosmic star formation rate (CSFR) is essential for using LGRBs
as cosmological probes. In this work, we collect a large sample composed
of 371 Swift LGRBs with known redshifts and prompt emission
properties. We first compare the rest-frame prompt properties of these bursts
in different redshift bins, finding negligible redshift evolution of
the luminosity of LGRBs with $L_{\mathrm{iso}}\gtrsim10^{51}\,\mathrm{erg\, s^{-1}}$
between $z\sim1$ and $z\sim4$. Then, by utilizing the CSFR obtained from
the large-scale cosmological hydrodynamical simulation,
the Illustris simulation, we calculate the cumulative redshift distribution
of LGRBs under different metallicity thresholds. After comparing with
our sample, we find that the predictions with a moderate threshold
between $0.3\,Z_{\odot}\leqslant Z_{\mathrm{th}}\leqslant1.0\,Z_{\odot}$
are consistent with the sample between redshift $0<z<3$, while at higher redshifts between $3<z<5$,
all metallicity thresholds fit the data well.
When changing to an empirical model based on observations,
the predictions show similar results as well.
After comparing with the metallicity distribution of the observed LGRB host galaxies between $0<z<1$,
we confirm that the production of LGRBs in galaxies with super-solar metallicity is suppressed.
Nevertheless, considering that a significant fraction of stars
are born in sub-solar metallicity environments at $z\gtrsim3$,
we suggest that, as a first approximation,
LGRBs can be used as direct tracers of the CSFR in this redshift range.
\end{abstract}


\section{Introduction}

Gamma-ray bursts (GRBs) have isotropic luminosities as high as $10^{54}\,\mathrm{erg\, s^{-1}}$,
making them detectable at the edge of the observable Universe. With
such an extremely high luminosity, GRBs are widely considered as
powerful probes of high-redshift galaxies once given the knowledge of their physical origin.
A phenomenological bimodal distribution of the observed
burst duration ($T_{90}$) with a rough separation around $2\,\mathrm{s}$ \citep{1993ApJ...413L.101K},
implies that these events are from two physically distinct progenitors.
That is, long GRBs (LGRBs) with $T_{90}>2\,\mathrm{s}$ are expected
to originate from the collapse of rapidly rotating massive stars \citep{1999ApJ...524..262M},
and short GRBs (SGRBs) with $T_{90}<2\,\mathrm{s}$ are from mergers
of compact stellar binaries \citep{1992ApJ...395L..83N}. Moreover,
the detection of associations of many LGRBs with core-collapse supernovae
\citep{2003Natur.423..847H,2006ARA&A..44..507W} as well as the preference
of LGRBs to be located in star-forming galaxies \citep[e.g.][]{2003A&A...400..499L,2009ApJ...691..182S}
provides strong observational support to the collapsar model of LGRBs.

Due to the link between LGRBs and the death of massive stars, the
redshift distribution of LGRBs can be used to study the cosmic star
formation rate (CSFR). Especially at very high redshifts where only the most
luminous galaxies are above the detection limit, LGRBs offer unique
opportunities as tracers of star formation in faint galaxies, independent
of galaxy luminosity and dust obscuration \citep[see][for recent reviews]{2015JHEAp...7...35S,2017RSOS....470304S}.
However, the way in which the LGRB rate traces the star formation
is still very uncertain. The core-collapse single progenitor model
for LGRBs predicts that low metallicity in the range of $Z<0.1-0.3\,Z_{\odot}$
is required for the retention of the angular momentum of the central
core to launch the jet \citep{2006ApJ...637..914W}, while binary
progenitor models typically imply a relatively modest metallicity
dependence \citep{2004MNRAS.348.1215I,2005ApJ...623..302F,2010MNRAS.406..840P}.
A number of past studies \citep[e.g.][]{2008ApJ...673L.119K,2008ApJ...683L...5Y,2010ApJ...711..495B,2012ApJ...744...95R}
have compared the LGRB redshift distribution to the CSFR and concluded that the LGRB cosmic efficiency
(i.e. the LGRB rate - CSFR ratio) increases rapidly with redshift over $z\lesssim4$, which
means that LGRBs are more frequent for a given CSFR at high redshifts.

Recent investigations of the observational data of host galaxies of
LGRBs make the situation a controversial topic. Several studies have
investigated the properties of LGRB hosts from unbiased and highly
complete samples, such as the BAT6 \citep{2012ApJ...749...68S}, TOUGH
\citep{2012ApJ...756..187H}, and SHOALS \citep{2016ApJ...817....7P}
samples. Based on these samples, studies focusing on the low redshifts
($z\lesssim1.5$) suggest that LGRBs are in general produced preferentially
in smaller, less massive and lower metallicity environments when compared
to typical star-forming galaxies, although the very low metallicity
thresholds expected from the single progenitor model are disfavored
\citep[e.g.][]{2015A&A...581A.125K,2015A&A...581A.102V,2016A&A...590A.129J,2016ApJ...817....8P}.
It is also interesting to note that, at low redshifts, LGRBs are dominated
by the so-called low-luminosity GRBs that may have different properties
from the high redshift GRBs \citep[e.g.][]{2007MNRAS.382L..21C,2007ApJ...662.1111L},
which produce a complex situation. The picture at higher redshifts
is much less clear because of the smaller size of LGRB samples. Some
studies \citep{2015ApJ...809...76G,2016ApJ...817....8P} claimed that
LGRBs can directly trace star formation at $z\gtrsim3$, which can
be explained by the fact that the average metallicity of star-forming
galaxies drops steeply from $z=0$ to $z\approx3$.
Meanwhile, \citet{2015ApJ...808...73S} found that the LGRBs in
a complete sample favor lower luminosity hosts at all comparable redshifts.

Alternatively, the properties of LGRB hosts have also been studied
by various theoretical models, including numerical simulations and
semi-analytical models. For example, \citet{2015MNRAS.446.4239E}
compared the results of high-resolution cosmological simulations with
the observed LGRB sample, suggesting that the LGRB cosmic efficiency is about
constant at $z>5$. Meanwhile, \citet{2017MNRAS.469.4921B} studied the
properties of LGRB host galaxies by using a simulated LGRB host sample
for $z<3$, indicating that LGRB progenitors favor the existence of
a metallicity threshold in the range of $0.3-0.6\,\mathrm{Z_{\odot}}$.
Assuming that LGRBs could originate from both a collapsar and a metal
independent channel, \citet{2015ApJ...802..103T} found a moderate
metallicity bias.

Although metallicity dependence has been expected to be the primary
factor affecting the LGRB cosmic efficiency, it is not necessarily the only
factor. Some other models have also been proposed to explain the nonconstant
LGRB cosmic efficiency, including the redshift-dependent LGRB luminosity
function \citep[e.g.][]{2011MNRAS.417.3025V,2016ApJ...820...66D}
and the redshift-dependent initial mass function of stars \citep[e.g.][]{2011ApJ...727L..34W}.
Moreover, there are several observational biases \citep[e.g.][]{2014ApJ...783...24L},
the most important effect of which is from the flux limit of the satellite.

In our previous study \citep{2013ApJ...772...42H}, we compared the
redshift distribution of Swift LGRBs from \citet{2012ApJ...744...95R}
with the predictions of several empirical star formation models, finding
a moderate metallicity bias. Moreover, based on a self-consistent star
formation model, we found that the difference between the CSFR
and the LGRB rate could be explained as a consequence of LGRBs
occurring in fainter galaxies. However, there are two main drawbacks
in this work. First, the significance of the results is limited
by the small number of LGRBs. Second, the adopted star formation
model is too simple to be able to describe the chemical evolution
in a self-consistent way.

In this paper, we gather redshifts and prompt emission properties
for a large sample of Swift LGRBs detected before the end of
2017 and examine whether the intrinsic properties of these bursts
in which we are interested are redshift-dependent. Then, by using
the CSFR obtained from the state-of-the-art cosmological hydrodynamic simulation of galaxy formation,
the Illustris simulation \citep{2014MNRAS.444.1518V,2014Natur.509..177V},
in which the star formation and metal
enrichment are treated in a self-consistent way, we calculate the
LGRB rate under different stellar metallicity thresholds. By comparing with our
sample, we investigate the connection between the LGRB
rate and star formation history to see if they can be used as an
unbiased tracer of star formation history. In order to account for
the effect of observational uncertainties, we also consider an empirical model of the
CSFR based on observations. As a further test, we also compare the expected metallicity
distribution of LGRB host galaxies with that of the observed host galaxies collected in the literature.

The paper is organized as follows. In Section 2, we describe our sample
of LGRBs with known redshifts and prompt emission properties
that we are interested in. In Section 3, the possible redshift evolution of
the LGRB luminosity distribution of our sample is checked. Section 4 outlines the
methods used to calculate the LGRB rate. The results of the comparisons
with observations are then shown in Section 5. Finally, conclusions
and discussion are presented in Section 6.

\section{LGRB sample}

Over the last decade, the Swift has detected more than 1000
GRBs with peak fluxes in the $15-150\,\mathrm{keV}$ energy band and
above the flux threshold of $\sim1\times10^{-8}\,\mathrm{erg\, cm^{-2}\, s^{-1}}$.
About one third of these bursts have redshift measurements by the multiwavelength
instruments on the satellite and follow-up observations through spectroscopy
and photometry of the GRB afterglow or the host galaxy.

We select our sample among all the Swift LGRBs
with known redshifts and prompt emission properties in the literature
before the end of 2017. Note that some events found by the ground analysis
but not by the satellite are removed, because their selections are done manually and lack systematicness.
Ultra-long GRBs marked by \citet{2016ApJ...829....7L} are also excluded,
since they may represent a distinct population of bursts that result
from different progenitors, such as the the tidal disruption events, magnetars, and
low-metallicity blue supergiants \citep{2013ApJ...766...30G,2014ApJ...781...13L}.
We also exclude GRBs that are marked as SGRBs in Greiner's online table
\footnote{\href{http://www.mpe.mpg.de/~jcg/grbgen.html}{http://www.mpe.mpg.de/$\sim$jcg/grbgen.html}}
where the classification also takes into account multiwavelength
criteria, although their durations are longer than $2\,\mathrm{s}$.
The sample constructed this way consists of 371 Swift LGRBs, as listed in
Table~\ref{tab1}. We note that their durations ($T_{90}$),
fluences within $15-150\,\mathrm{keV}$ ($S$), and peak photon fluxes
($P$) are taken from the online Swift/BAT GRB catalog
\footnote{\href{http://swift.gsfc.nasa.gov/results/batgrbcat/index.html}{http://swift.gsfc.nasa.gov/results/batgrbcat/index.html}} \citep{2016ApJ...829....7L}.

\begin{figure}
\begin{center}
\includegraphics[angle=0,width=0.8\textwidth]{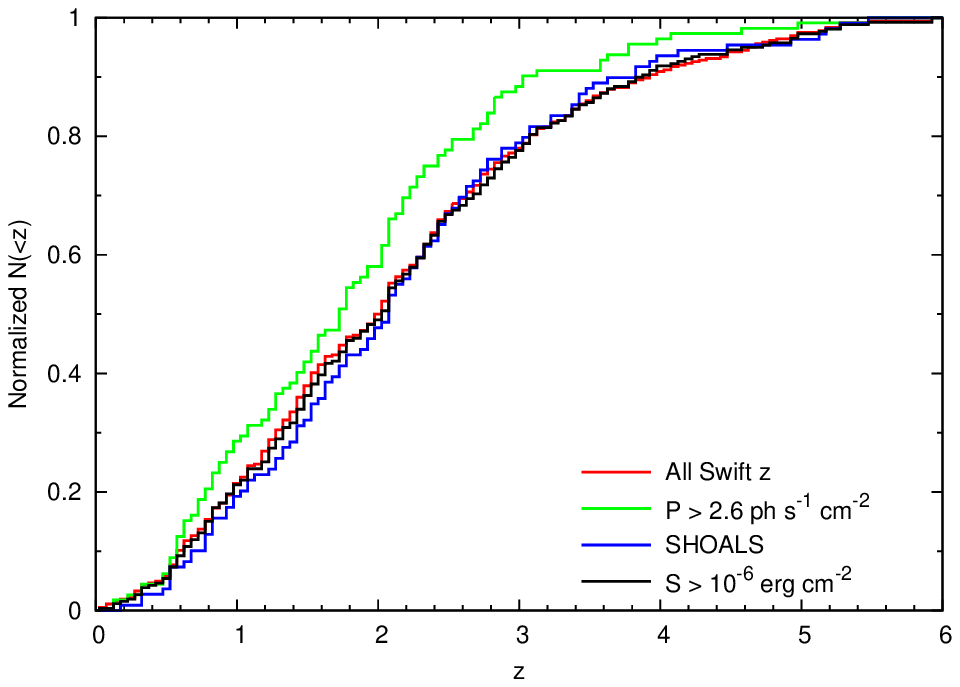}
\end{center}
\caption{Cumulative redshift distribution of our sample with
a fluence cut ($S_{\mathrm{15-150keV}}>10^{-6}\mathrm{erg\, cm^{-2}}$)
over $0<z<6$ compared to other LGRB samples, including the overall
distribution of 371 Swift LGRBs with known redshifts, the SHOALS
sample and so on.\label{fig1}}
\end{figure}

One major selection effect for the observed redshift distribution
of LGRBs is from the survey sensitivity of Swift's detectors.
However, the trigger criteria of Swift/BAT
are very complicated, and there are two stages to this
criteria. The ``rate trigger'' stage adopts over 500 ``rate trigger'' criteria,
according to which a brighter burst has a higher trigger probability.
In addition, an image will be generated for further confirmation and localization,
the trigger of which also depends on the burst duration. A single explicit cut on prompt
emission properties is commonly used for the purpose of measuring intrinsic redshift distributions. For
example, \citet{2012ApJ...749...68S} adopted a peak flux cut of $P\geqslant2.6\,\mathrm{ph\, s^{-1}\, cm^{-2}}$,
which corresponds to a peak flux that is $\sim6$ times larger than
the peak flux threshold of Swift/BAT.
Nevertheless, the peak flux cut would affect the redshift distribution by
removing more high redshift bursts with low-flux than that at low redshifts,
while a fluence cut is suggested to have less influence on the redshift distribution \citep{2016ApJ...817....7P}.
Hence, we choose to adopt a fluence cut of $S_{15-150\,\mathrm{keV}}\geqslant 1.0\times10^{-6}\,\mathrm{erg\, cm^{-2}}$,
same as in \citet{2016ApJ...817....7P}.
This further criterion selects 261 LGRBs, with a completeness in the redshift of 44 percent (261/597).
Although the completeness of the sample with
$P\geqslant2.6\,\mathrm{ph\, s^{-1}\, cm^{-2}}$ in the redshift is 48 percent (112/235),
which is slightly higher than the sample with the fluence cut, but the fluence cut criterion could
provide a sample that has a much larger size. The
cumulative redshift distribution of the fluence cut sample is shown in Figure~\ref{fig1}. As can be seen, the
redshift distribution of this sample is very similar to that of all
known-redshift Swift LGRBs. In contrast,
the redshift distribution of LGRBs with the peak flux cut shows an
obvious excess of LGRBs at lower redshifts.
The Anderson-Darling (A-D) k-sample test for consistency
gives a $p=0.02$ between this distribution with the peak flux cut and that with the fluence cut. For comparison,
we also show the distribution of the complete SHOALS \citep{2016ApJ...817....7P} sample in Figure~\ref{fig1}.
It is worth stressing that although the completeness of our sample is not as high as
the SHOALS sample, they show similar consistency.

\begin{figure}
\begin{center}
\includegraphics[angle=0,width=0.8\textwidth]{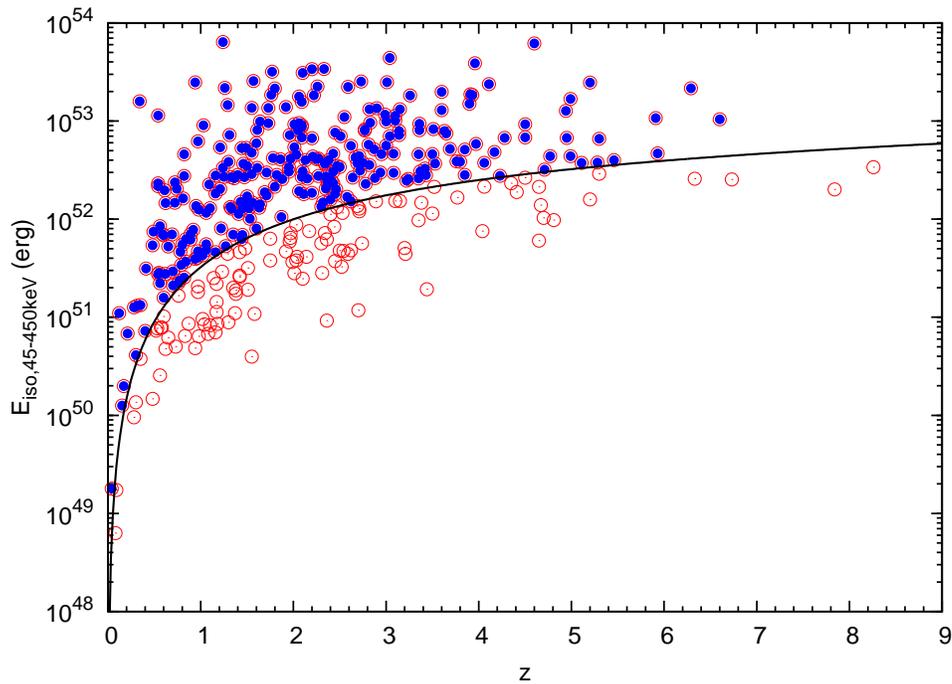}
\end{center}
\caption{Isotropic energy vs. redshift distribution of 371 Swift
LGRBs. The blue filled circles represent the LGRBs with a fluence cut, while the red empty circles
represent all LGRBs with known redshifts.
The solid line represents the energy threshold
with $S_{\mathrm{lim},15-150\,\mathrm{keV}}=10^{-6}\,\mathrm{erg\, cm^{-2}}$.\label{fig2}}
\end{figure}

\begin{figure}
\begin{center}
\includegraphics[angle=0,width=0.8\textwidth]{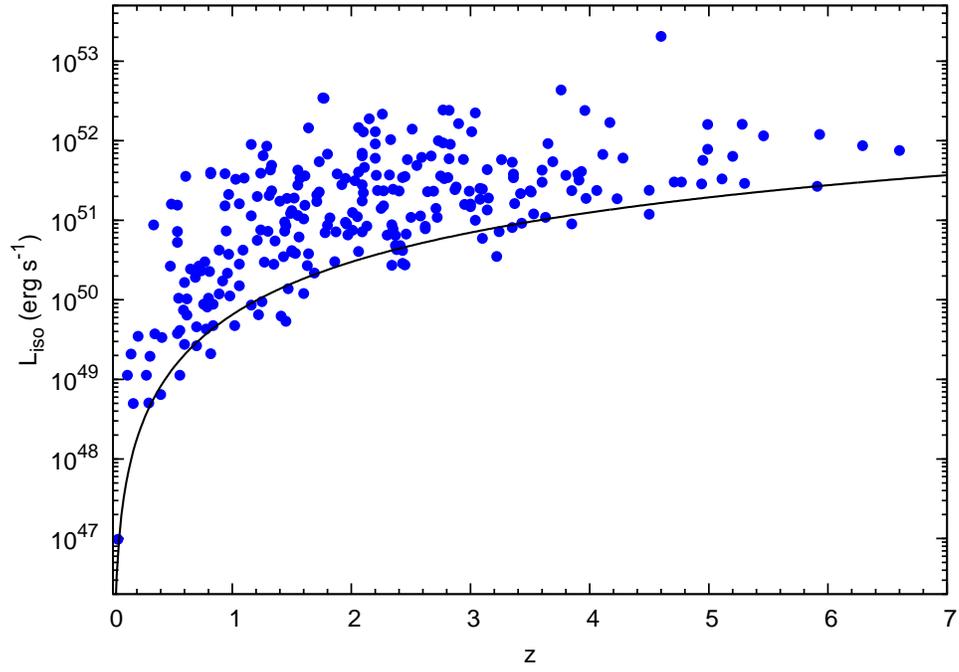}
\end{center}
\caption{Isotropic luminosity vs. redshift distribution of LGRBs with the fluence cut.
The solid line represents the detection threshold of Swift, which has a flux limit of
$\sim F_{\mathrm{lim}}=1.0\times 10^{-8}\,\mathrm{erg\, cm^{-2}\, s^{-1}}$.\label{fig3}}
\end{figure}

From the redshift $z$ and the fluence $S$, integrated over the observed
Swift band (15-150 keV), we
calculate the intrinsic isotropic energy $E_{\mathrm{iso}}$, following the
standard relation:
\begin{equation}
E_{\mathrm{iso,45-450\, keV}}=\frac{4\pi d_{L}^{2}S_{\mathrm{15-150\, keV}}k(z)}{1+z},
\end{equation}
where $d_{L}$ is the luminosity distance and $k(z)$ is the \emph{k}-correction
for the fluence transferred from the observed detector band to its
rest-frame. Following \citet{2016ApJ...817....7P}, in order to reduce
the uncertainty due to Swift's narrow bandpass, we only calculate
the rest-frame energy range of 45-450 keV, which would make the \emph{k}-correction
much smaller and more reliable. In particular, its value is $k(z)=[(1+z)/(1+2)]^{-0.5}$.
Figure~\ref{fig2} shows the isotropic energy-redshift distribution
of our sample, in which the solid line indicates the energy threshold
with $S_{\mathrm{lim},15-150\,\mathrm{keV}}=1.0\times10^{-6}\,\mathrm{erg\, cm^{-2}}$.
Then, the isotropic luminosity $L_{\mathrm{iso}}$ is estimated as
\begin{equation}
L_{\mathrm{iso}}=\frac{E_{\mathrm{iso,45-450\, keV}}}{T_{90}/(1+z)}.
\end{equation}
In Figure~\ref{fig3}, the isotropic luminosity distribution is shown for bursts with the fluence cut.
The solid line indicates the detection threshold of Swift, which has a flux limit of
$\sim F_{\mathrm{lim}}=1.0\times 10^{-8}\,\mathrm{erg\, cm^{-2}\, s^{-1}}$.

\section{Check for the luminosity evolution}

\begin{figure}
\begin{center}
\includegraphics[angle=0,width=0.8\textwidth]{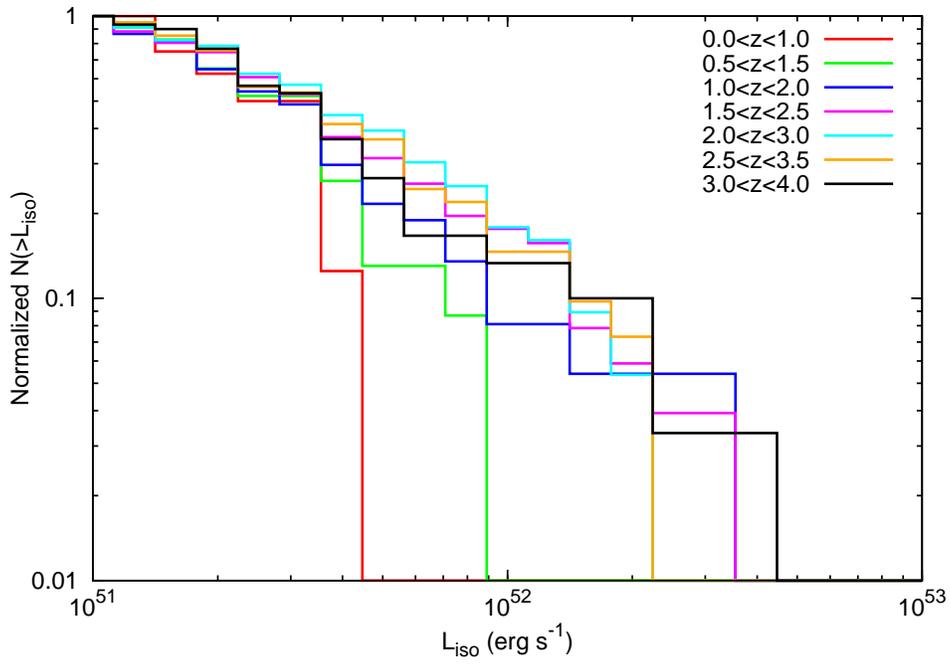}
\end{center}
\caption{Cumulative luminosity distributions of LGRBs with $L_{\mathrm{iso}}>10^{51}\,\mathrm{erg\, s^{-1}}$
in seven different redshift bins.\label{fig4}}
\end{figure}

\begin{figure}
\begin{center}
\includegraphics[angle=0,width=0.8\textwidth]{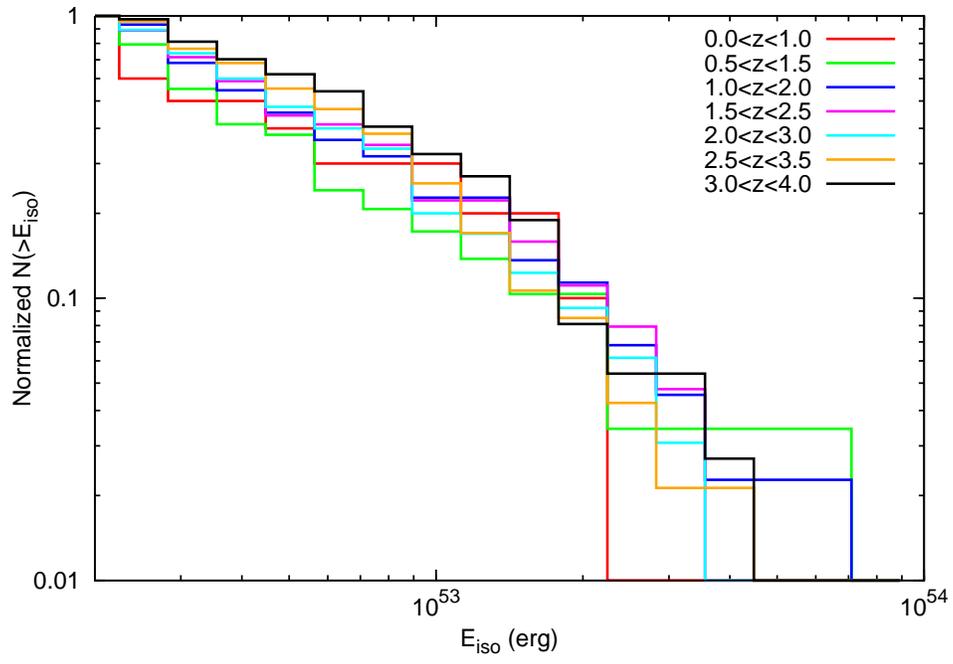}
\end{center}
\caption{Same as Figure~\ref{fig4}, but for the isotropic energy
of LGRBs with $E_{\mathrm{iso}}>2\times10^{52}\,\mathrm{erg}$.\label{fig5}}
\end{figure}

In addition to the metallicity threshold, the cosmic evolution of the LGRB
luminosity could be one of the explanations for the nonconstant LGRB cosmic
efficiency. Here, with our relatively large sample, we first
check whether such a cosmic evolution of the LGRB luminosity exists using a simple method.
We divide our LGRB sample into seven redshift bins
as $0<z<1$, $0.5<z<1.5$, $1<z<2$, $1.5<z<2.5$, $2<z<3$, $2.5<z<3.5$, and $3.0<z<4.0$.
The adjacent bins are adopted to overlap with each other just to obtain
a sufficiently large LGRB number for each redshift bin.
In Figure~\ref{fig4}, as a visualization of the redshift evolution,
the cumulative luminosity distributions of LGRBs with $L_{\mathrm{iso}}>10^{51}\,\mathrm{erg\, s^{-1}}$
in various redshift bins are displayed. As can be seen,
the two lowest-redshift bins show an obvious lack of medium-
to high-luminosity LGRBs, which may be due to their low
event rate and the small observable volume at low redshifts.
On the other hand, the remaining five curves between $1<z<4$ display similar consistency.
The A-D test gives a $p=0.83$ that these five distributions are drawn from the same parent distribution,
which implies that there is little redshift evolution of the LGRB luminosity, at least
between $1<z<4$. We also plot the cumulative distributions
of isotropic energy in different redshift bins in Figure~\ref{fig5}.
These distributions show similar characteristics to that of the LGRB
luminosity. Therefore, it is enough
to assume a non-evolving luminosity function of LGRBs with $L_{\mathrm{iso}}\gtrsim10^{51}\,\mathrm{erg\, s^{-1}}$
for our purpose in this paper.
We leave a more thorough investigation on the redshift evolution of
the luminosity distribution to a future work.

\section{LGRB formation rate}

In this section, we estimate the rate of LGRB formation as a function
of redshift. LGRBs are believed to result from the collapse
of massive stars, making them good tracers of the star formation rate after
taking into account the conditions necessary to affect the LGRB cosmic efficiency.
Our estimations thus begin with the CSFR. In this
paper, the CSFR is obtained from
the Illustris simulation \citep{2014MNRAS.444.1518V,2014MNRAS.445..175G},
which is a large-scale cosmological hydrodynamical simulation of galaxy formation in a volume of
$(106.5\,\mathrm{Mpc})^{3}$ with a dark mass resolution of $6.26\times 10^{6}\,\mathrm{M_{\odot}}$ and
a baryonic mass resolution of $1.26\times 10^{6}\,\mathrm{M_{\odot}}$.
The dynamics of dark matter and gas are computed using the moving-mesh code AREPO \citep{2010MNRAS.401..791S}.
The simulation adopted the following cosmological parameters: $\Omega_{m}=0.2726$,
$\Omega_{\Lambda}=0.7274$, $\Omega_{b}=0.0456$, $\sigma_{8}=0.809$, $n_s=0.963$ and
$h=0.704$, consistent with the Wilkinson Microwave Anisotropy Probe (WMAP)-9 results \citep{2013ApJS..208...19H}.
In addition, the subgrid physical processes for galaxy formation employed by Illustris include gas
cooling (primordial and metal-line cooling),
stellar evolution and feedback processes, gas recycling, metal enrichment based on nine elements,
and supermassive black hole growth and related active galactic nucleus (AGN) feedback processes in various modes.
For full details of the models and the parameter selection see \citet{2013MNRAS.436.3031V} and \citet{2014MNRAS.438.1985T}.
The simulation starts at redshift $z=127$ and evolves to the present day ($z=0$).
Many of the key observables of the local Universe, such as the galaxy stellar mass and luminosity functions,
baryon conversion efficiency and the morphology of galaxies, are reasonably well reproduced by the simulation,
although not every observed property is matched precisely.
There are still some discrepancies, such as the quenching of massive
galaxies and the age distribution of low-mass galaxies \citep{2014MNRAS.444.1518V}.
The subgrid models of the simulation still need further improvements.
Nevertheless, the simulation could provide us with useful
insights on the physics of the Universe especially at high redshifts where direct observations are generally lacking.

\begin{figure}
\begin{center}
\includegraphics[angle=0,width=0.8\textwidth]{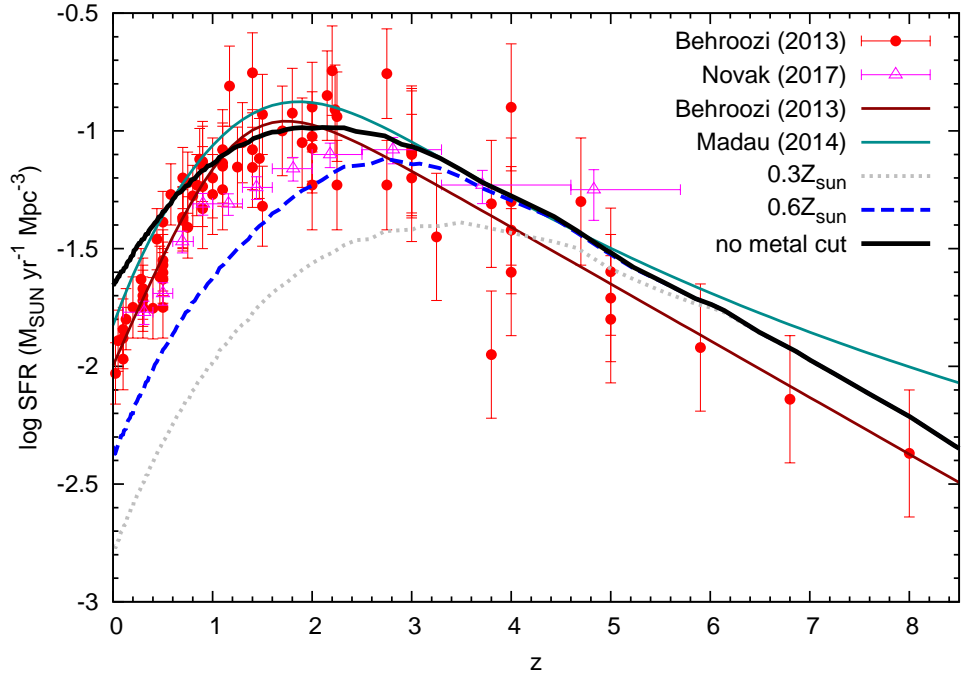}
\end{center}
\caption{Cosmic star formation rate from the Illustris
simulation and estimations with galaxy stellar metallicities below a certain value:
0.3 and 0.6~$Z_{\odot}$. The observational data are taken from UV+IR measurements \citep{2013ApJ...770...57B}
and radio measurements \citep{2017A&A...602A...5N}. As a reference, we also plot the
empirical fit of \citet{2013ApJ...770...57B} and the CSFR
of \citet{2014ARA&A..52..415M}.\label{fig6}}
\end{figure}

The predicted global CSFR from the Illustris simulation is also in good agreement with the observations
\citep{2013ApJ...770...57B,2017A&A...602A...5N} up to $z\sim8$, as shown in Figure~\ref{fig6}.
The slight excess of the simulated star formation at lower redshifts seems to be the result of insufficient AGN feedback to quench star
formation at these times, as indicated by \citet{2014MNRAS.444.1518V}.
The contributions to the CSFR from stellar populations with metallicities below different thresholds ($Z_{\mathrm{th}}=0.3$ and $0.6\,Z_{\odot}$, for example) are also shown,
which can be easily used to deduce the preference of LGRBs for low-metallicity progenitors.
As can be seen, a large fraction of new born stars have metallicities above the threshold of $Z_{\mathrm{th}}=0.6\,Z_{\odot}$ at lower redshifts,
while at $z\gtrsim3$ most stars are born with metallicities below this value.
In order to account for the uncertainty in the star formation history, we also plot the CSFR fits from \citet{2013ApJ...770...57B}
and \citet{2014ARA&A..52..415M}, respectively, in Figure~\ref{fig6}.

If we assume that the progenitors of LGRBs are stars massive enough
to form a black hole, the LGRB rate should follow the CSFR in a unbiased
way. Then, the intrinsic cosmic LGRB rate can be related to the CSFR as
\begin{equation}
\dot{n}_{\mathrm{LGRB}}(z)\propto\dot{\rho}_{\mathrm{SFR}}(z).
\end{equation}
Note that the time delayed between the formation of LGRBs and the
progenitor stars is negligible, since massive stars are short-lived
objects.

Adopting the CSFR from the Illustris simulation, a low-metallicity
preference for LGRB progenitors could be simply described by considering
only the new born stars with the metallicity below a given
threshold ($Z_{\mathrm{th}}$). Then, the intrinsic cosmic LGRB rate
for a selected $Z_{\mathrm{th}}$ is
\begin{equation}
\dot{n}_{\mathrm{LGRB}}(z)\propto\dot{\rho}_{\mathrm{SFR}}(z,\, Z_{\mathrm{th}}).
\end{equation}

As a reference for the observational uncertainty, we also derive the intrinsic LGRB rate
from the CSFR fit of \citet{2013ApJ...770...57B} as
\begin{equation}
\dot{n}_{\mathrm{LGRB}}(z)\propto\Psi(z,\, Z_{\mathrm{th}})\dot{\rho}_{\mathrm{SFR}}(z),
\end{equation}
where $\Psi(z,\, Z_{\mathrm{th}})$ is the fraction of star formation
occurring in galaxies with a metallicity below $Z_{\mathrm{th}}$,
which is described by the following expression \citep{2012ApJ...744...95R}:
\begin{equation}
\Psi(z,\, Z_{\mathrm{th}})=\frac{\int_{0}^{M_{\star,crit}(z,Z_{\mathrm{th}})}\mathrm{SFR}(M_{\star},\, z)\phi(M_{\star},\, z)dM_{\star}}{\int_{0}^{\infty}\mathrm{SFR}(M_{\star},\, z)\phi(M_{\star},\, z)dM_{\star}},\label{eq:6}
\end{equation}
where $\mathrm{SFR}(M_{\star},\, z)$ is the SFR-stellar mass relation,
$\phi(M_{\star},\, z)$ is the galaxy stellar mass function, and $M_{\star,crit}(z,Z_{\mathrm{th}})$
is the critical galaxy mass related to a metallicity threshold $Z_{\mathrm{th}}$,
which can be expressed by the so-called redshift-dependent mass-metallicity
($M$-$Z$) relation (\citealt{2005ApJ...635..260S}, on the scale
of \citealt{2004ApJ...617..240K})%
\footnote{The oxygen abundance 12+log{[}O/H{]} is related to the metallicity
$Z/Z_{\odot}$ by the scaling $\mathrm{12+log[O/H]}=\log(Z/Z_{\odot})+8.69$
\citep{2001ApJ...556L..63A} throughout this paper.}
\begin{eqnarray}
12+\log[\mathrm{O/H}] & = & -7.5903+2.5315\log M_{\star}\nonumber \\
 &  & -0.09649\log^{2}M_{\star}\nonumber \\
 &  & +5.1733\log t_{\mathrm{u}}-0.3944\log^{2}t_{\mathrm{u}}\nonumber \\
 &  & -0.403\log t_{\mathrm{u}}\log M_{\star},
\end{eqnarray}
where $t_{\mathrm{u}}$ is the age of the Universe at redshift $z$ in Gyr
and $M_{\star}$ is the galaxy stellar mass in $M_{\odot}$. \citet{2016ApJ...817..118T}
parameterized the redshift-dependent SFR-stellar mass relation at $0.5<z<4$
using data from the FourStar Galaxy Evolution Survey, based on the
formula of \citet{2015ApJ...801...80L}:
\begin{equation}
\log(\mathrm{SFR}(M_{\star},\, z))=s_{0}-\log\left[1+\left(\frac{M_{\star}}{M_{0}}\right)^{-\gamma}\right],
\end{equation}
where the best-fit parameters $s_{0}$ in $\log(M_{\odot}/\mathrm{yr})$
and $M_{0}$ in $M_{\odot}$ for star-forming galaxies are
\begin{eqnarray*}
s_{0} & = & 0.448+1.220z-0.174z^{2}\\
\log(M_{0}) & = & 9.458+0.865z-0.132z^{2}\\
\gamma & = & 1.091.
\end{eqnarray*}
For the redshift-evolving stellar mass function, we adopt the parameterization
derived by \citet{2008ApJ...680...41D}:
\begin{equation}
\phi(M_{\star},\, z)dM_{\star}=\phi_{\star}\left(\frac{M_{\star}}{M_{1}}\right)^{\gamma}\exp\left(-\frac{M_{\star}}{M_{1}}\right)\frac{dM_{\star}}{M_{1}},
\end{equation}
where the parametric functions obey
\begin{eqnarray*}
\phi_{\star}(z) & \approx & 0.003(1+z)^{-1.07}\,\mathrm{Mpc^{-3}dex^{-1}}\\
\log[M_{1}/M_{\odot}](z) & \approx & 11.35-0.22\ln(1+z)\\
\gamma & \approx & -1.3.
\end{eqnarray*}

\begin{figure}
\begin{center}
\includegraphics[angle=0,width=0.8\textwidth]{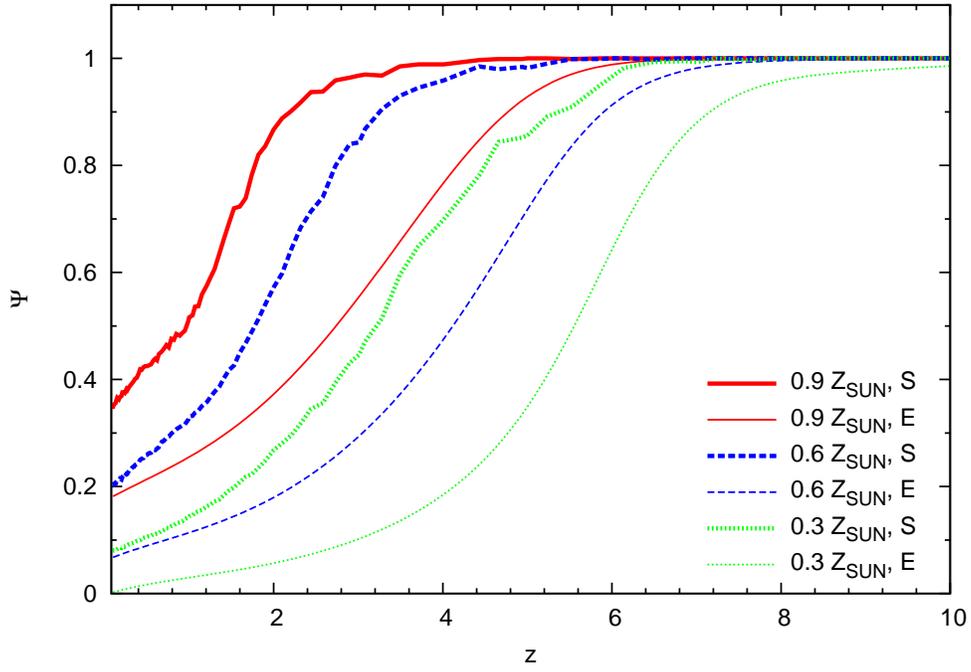}
\end{center}
\caption{Fraction of star formation occurring in galaxies below
a given metallicity threshold of $Z/Z_{\odot}$ as a function of redshift (denoted as E).
As a reference, we also plot the corresponding results from the Illustris simulation (denoted as S).\label{fig7}}
\end{figure}

In Figure~\ref{fig7}, we show the results of Equation (\ref{eq:6})
with different metallicity thresholds: $Z_{\mathrm{th}}=0.3$, 0.6 and $0.9\,Z_{\odot}$.
The results from the Illustris simulation are also plotted as a comparison.
As can be seen, the empirical model shows that more than $80\%$ of stars are formed in galaxies above $0.9\,Z_{\odot}$ around $z=0$,
while the simulation predicts a somehow lower fraction ($\sim60\%$) at the same redshift.
The discrepancy between the simulation and the empirical model could come from various uncertainties.
For instance, the empirical models are heavily dependent on the adopted functional form and observational data.
Different empirical models can suggest very different imprints, as seen in our previous work \citep{2013ApJ...772...42H}.
Also note that the empirical relations, such as the $M$-$Z$ relation of \citet{2005ApJ...635..260S}
are mainly based on the low redshift data,
the validity of which at high redshifts needs to be tested and improved with future measurements.
On the other hand, \citet{2017MNRAS.469.4921B} indicated that the $M$-$Z$ relation of the Illustris simulation presents,
more or less, some differences with observations, although it could globally reproduce a large number of key LGRB host properties.
However, a thorough investigation about this discrepancy is beyond the reach of this paper.

The expected redshift distribution of LGRBs is given by
\begin{equation}
\frac{dN}{dz}=A\dot{n}_{\mathrm{LGRB}}(z,\, Z_{\mathrm{th}})\frac{dV/dz}{1+z},
\end{equation}
where the constant $A$ depends on the observing time, the sky coverage,
and so on. The comoving volume element $dV/dz$ is calculated by
\begin{equation}
\frac{dV}{dz}=\frac{4\pi cd_{\mathrm{L}}^{2}}{1+z}\left|\frac{dt}{dz}\right|,
\end{equation}
where $d_{\mathrm{L}}$ is the luminosity distance and $dt/dz$
is given by \citep{2010MNRAS.401.1924P}
\begin{equation}
\frac{dt}{dz}=\frac{9.78h^{-1}\mathrm{Gyr}}{(1+z)\sqrt{\Omega_{\Lambda}+\Omega_{\mathrm{m}}(1+z)^{3}}}.
\end{equation}
Hence, the number
of expected LGRBs in the redshift between $z_{1}$ and $z_{2}$ is
\begin{equation}
N(z_{1},\, z_{2})=A\int_{z_{1}}^{z_{2}}\dot{n}_{\mathrm{LGRB}}(z,\, Z_{\mathrm{th}})\frac{dV/dz}{1+z}dz.\label{eq:13}
\end{equation}
To remove the constant $A$, the cumulative redshift distribution
of LGRBs can be expressed as
\begin{equation}
N(z_{1}|<z|z_{\mathrm{max}})=\frac{N(z_{1},\, z)}{N(z_{1},\, z_{\mathrm{max}})}.
\end{equation}

\section{Comparison with observations}

\subsection{Cumulative redshift distribution of LGRBs}

Considering that the low-metallicity galaxies increasingly dominate the contribution
to the CSFR at higher redshifts, even if a metallicity
bias is present in the LGRB progenitors, the expected correction from
the LGRB rate to the CSFR at high redshifts would be limited. In fact, the
investigation of LGRB host galaxies over $3<z<5$ suggests that LGRBs
are an unbiased star formation tracer from $z\approx3$ out to the
highest redshift \citep{2015ApJ...809...76G}. Thus, we divide our
LGRB sample into two components: a low-redshift sample in the range of
$0<z<3$, and a high-redshift sample in the range $3<z<5$. In order
to remove low-luminosity bursts that could not be detected at high redshifts,
which is due to the selection effects from the Swift threshold,
we choose a luminosity cut of $L_{\mathrm{iso}}>0.7\times10^{51}\,\mathrm{erg\, s^{-1}}$
for the low-redshift sample, which has 120 LGRBs left, while
the luminosity cut is set to $L_{\mathrm{iso}}>2\times10^{51}\,\mathrm{erg\, s^{-1}}$
for the high-redshift sample, which includes 35 LGRBs.

\begin{figure}
\begin{center}
\includegraphics[angle=0,width=0.8\textwidth]{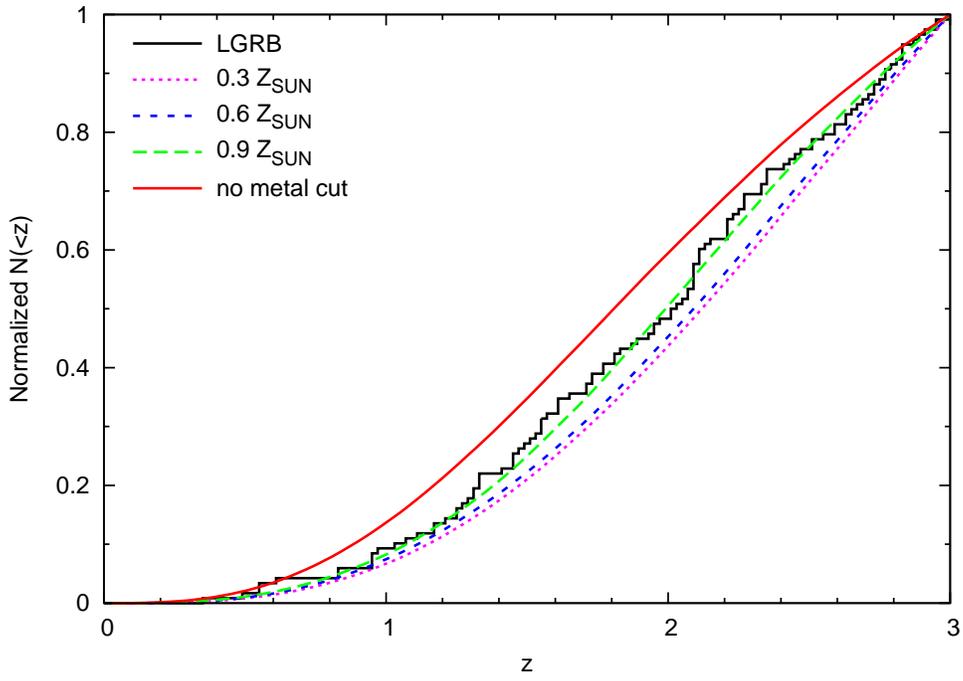}
\end{center}
\caption{Cumulative redshift distribution of 120 Swift
LGRBs with $L_{\mathrm{iso}}>0.7\times10^{51}\,\mathrm{erg\, s^{-1}}$
between $0<z<3$. The expected distributions are calculated using the
CSFR from the Illustris simulation, for different metallicity thresholds.\label{fig8}}
\end{figure}

\begin{figure}
\begin{center}
\includegraphics[angle=0,width=0.8\textwidth]{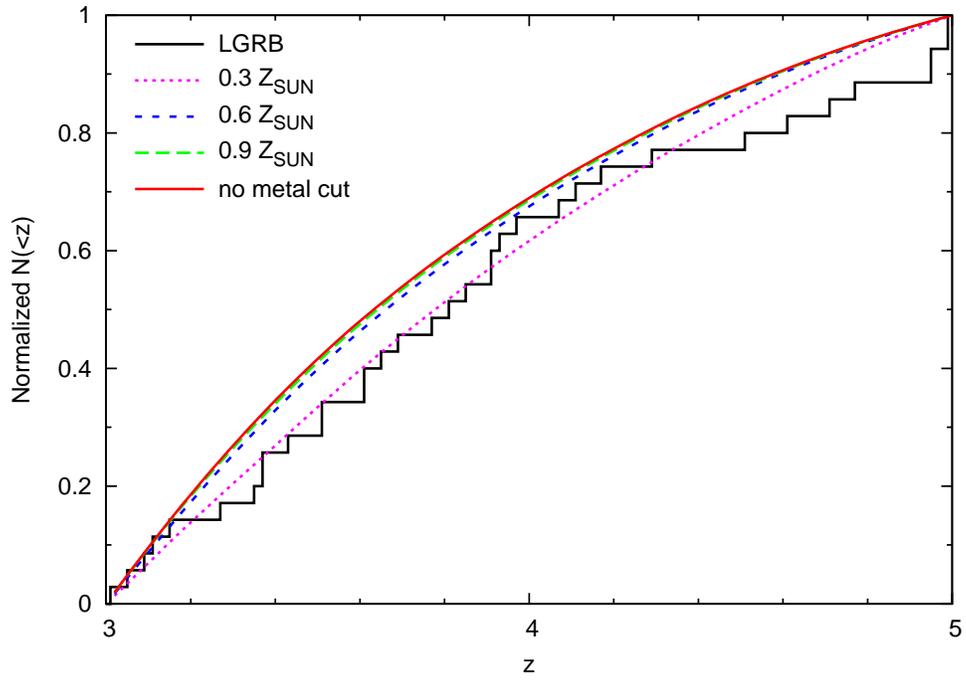}
\end{center}
\caption{Cumulative redshift distribution of 35 Swift
LGRBs with $L_{\mathrm{iso}}>2\times10^{51}\,\mathrm{erg\, s^{-1}}$
between $3<z<5$. The expected distributions are inferred from the Illustris simulation.\label{fig9}}
\end{figure}

\begin{figure}
\begin{center}
\includegraphics[angle=0,width=0.8\textwidth]{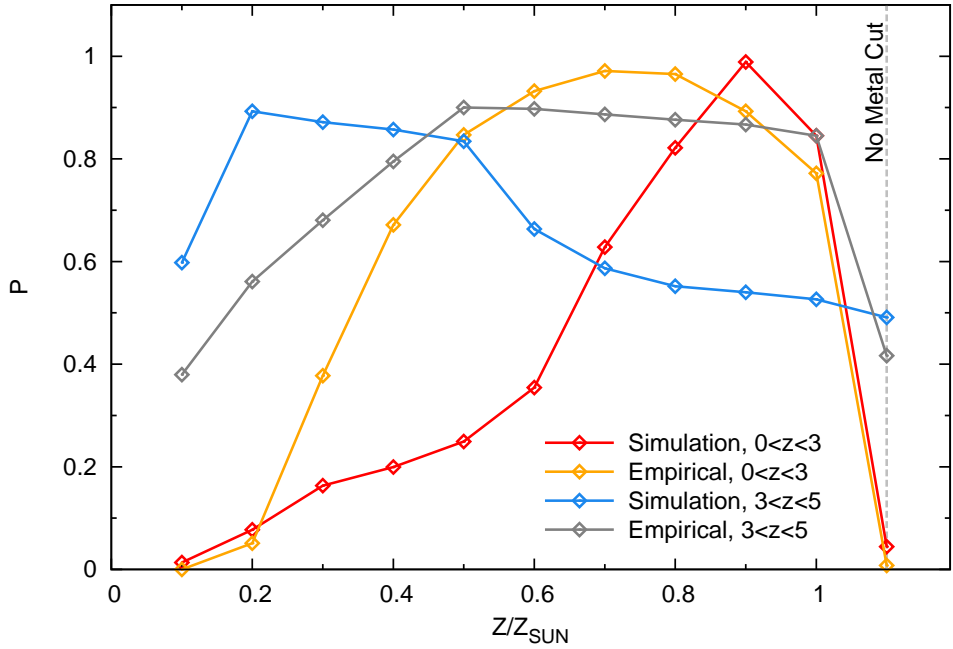}
\end{center}
\caption{Results of the K-S tests comparing different expected cumulative redshift distributions with
the observed LGRB samples. The $p$-values of consistency are calculated by assuming
different metallicity thresholds for LGRB production: $Z_{\mathrm{th}}=0.1$, 0.2, 0.3, \ldots,
$1.0\,Z_{\odot}$, as well as the null hypothesis that no metallicity preference exists.
Note that the rightmost points of these lines are for the cases with no metal cutoff.
\label{fig10}}
\end{figure}

The cumulative redshift distribution of the low-redshift sample between
$0<z<3$ is shown in Figure~\ref{fig8}. The solid red line is the
expected LGRB distribution based on the Illustris CSFR with no metallicity threshold.
The LGRB distributions inferred by adopting $Z_{\mathrm{th}}=0.3$, 0.6, and $0.9\,Z_{\odot}$, respectively, are shown
in Figure~\ref{fig8} for instance.
Whether a model distribution is consistent with that of the observed sample can be evaluated by the one-sample
Kolmogorov-Smirnov (K-S) test.
After considering the given metallicity thresholds: $Z_{\mathrm{th}}=0.1$, 0.2, 0.3, \ldots,
$1.0\,Z_{\odot}$, as well as the null hypothesis that there is no metallicity preference,
we find that all metallicity thresholds between
$Z_{\mathrm{th}}=0.3-1.0\,Z_{\odot}$ could produce K-S test $p>0.1$.
The maximum probability occurs at
$Z_{\mathrm{th}}=0.9\,Z_{\odot}$ (K-S test $p=0.99$).
Cases with no metallicity threshold or with an extreme threshold of
$Z_{\mathrm{th}}\leqslant0.2\,Z_{\odot}$ are disfavored (K-S test $p<0.07$).
This confirms the presence of a host galaxy metallicity threshold close to the solar value,
above which LGRBs are suppressed \citep{2012ApJ...744...95R,2014ApJS..213...15W,2016ApJ...817....8P}.
On the other hand, for the high-redshift sample between $3<z<5$,
we find that all metallicity thresholds produce K-S test $p>0.49$ (Figure~\ref{fig9}).
This is somewhat not surprising, since at $z>3$, most of the star formation
occurs in galaxies of $Z\lesssim0.6\,Z_{\odot}$ (see Figure~\ref{fig6}).
This may imply that LGRBs could be used as unbiased tracers of
star formation at high redshifts, in accordance with the analysis of host galaxies over $3<z<5$ by \citet{2015ApJ...809...76G}.
Nevertheless, we caution that this conclusion is only tentative due to
the small number of LGRBs in this redshift interval.
The results of the K-S tests are summarized in Figure~\ref{fig10}.

\begin{figure}
\begin{center}
\includegraphics[angle=0,width=0.8\textwidth]{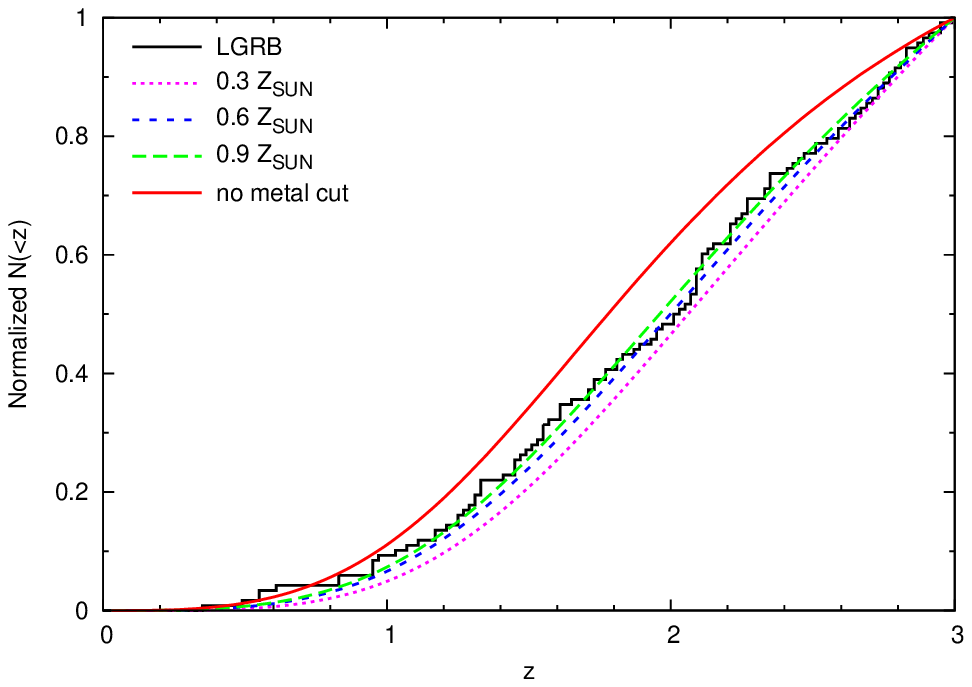}
\end{center}
\caption{Same as Figure~\ref{fig8}, but for the empirical model.\label{fig11}}
\end{figure}

\begin{figure}
\begin{center}
\includegraphics[angle=0,width=0.8\textwidth]{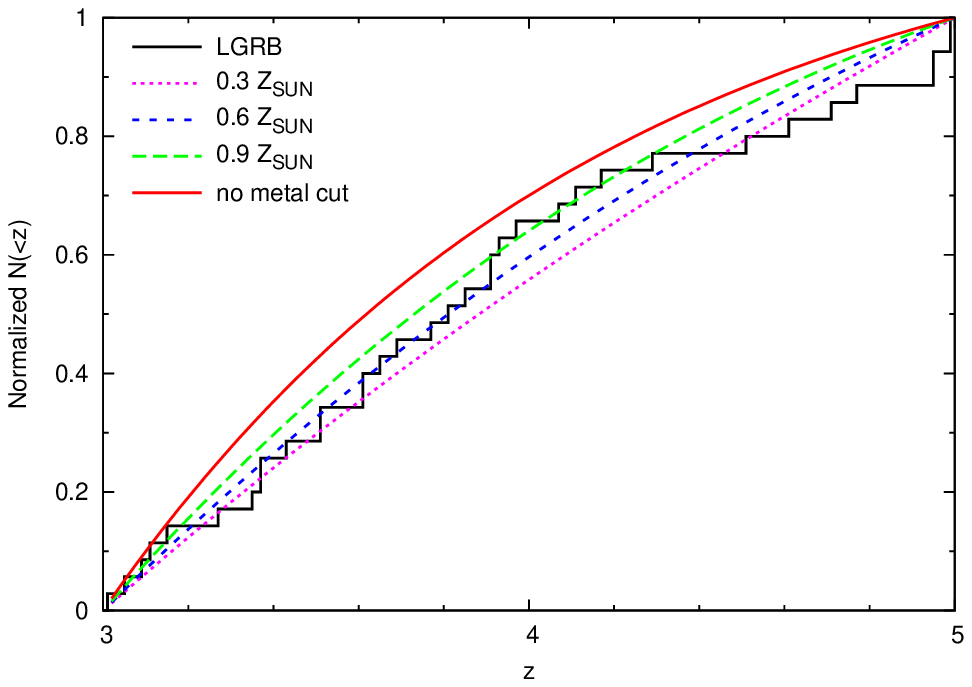}
\end{center}
\caption{Same as Figure~\ref{fig9}, but for the empirical model.\label{fig12}}
\end{figure}

For comparison, we also calculate the expected cumulative redshift distribution
of LGRBs using the empirical CSFR from \citet{2013ApJ...770...57B} together with
Equation (\ref{eq:6}), for different metallicity threshold values.
The results are shown in Figures~\ref{fig11}
and~\ref{fig12}, for low and high redshifts, respectively (see also Figure~\ref{fig10} for the K-S test results).
Similar to the results from the Illustris simulation,
we find that the region where K-S test $p>0.1$ contains metallicity thresholds of
$0.3\,Z_{\odot}\leqslant Z_{\mathrm{th}}\leqslant 1.0\,Z_{\odot}$ and
the maximum $p$-value occurs at $Z_{\mathrm{th}}=0.7\,Z_{\odot}$ at $0<z<3$.
For the redshift range between 3 and 5,
all cases are consistent with the data (K-S test $p>0.38$), which is also in accordance with the
results from the Illustris simulation.
It is interesting to note that,
while the cumulative LGRB distribution for the metallicity threshold of
$Z_{\mathrm{th}}=1.0\,Z_{\odot}$ shows a good agreement with the observations,
the result from a lower threshold of $Z_{\mathrm{th}}=0.3\,Z_{\odot}$ presents similar consistency as well.
However, the absolute metallicity distributions of LGRB progenitors would be quite different
between these two thresholds.
Therefore, a further test is necessary to understand their intrinsic metallicity preference, which will be
executed in the next section.

\subsection{Metallicity distribution of LGRB host galaxies}

\begin{figure}
\begin{center}
\includegraphics[angle=0,width=0.8\textwidth]{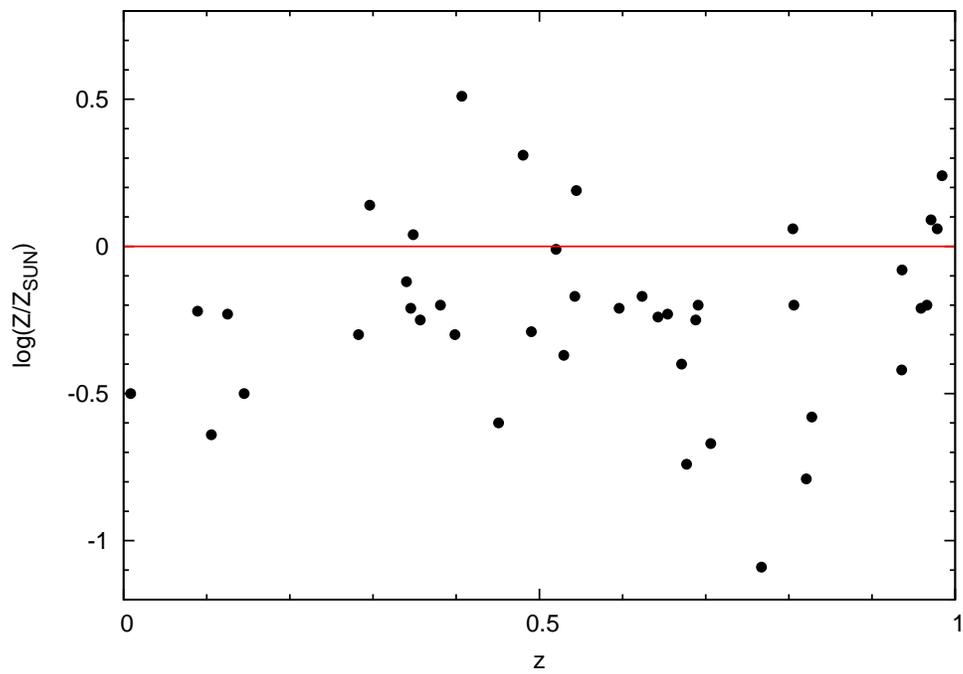}
\end{center}
\caption{Metallicity vs. redshift distribution of 42 LGRB host galaxies between $0<z<1$.
The red solid line represents the solar value $Z_{\odot}$.\label{fig13}}
\end{figure}

To further understand the effects of metallicity on the comparison of cumulative redshift distributions,
here we investigate the metallicity distribution of LGRB host galaxies.
Using Equation (\ref{eq:13}), we can simply estimate
the number of stars being produced as a function of stellar metallicity.
This result can then be compared with the metallicity distribution of LGRB host galaxies from the observations.
For comparison, we collect a sample of LGRB host galaxies with metallicities determined by direct observations
in the literature, including \citet{2015A&A...581A.125K}, \citet{2016A&A...590A.129J}, and GRB Host Studies (GHostS)\footnote{\href{http://www.grbhosts.org}{http://www.grbhosts.org}}.
As there are plenty unknown issues relating to the cosmic evolution, we would restrict our host sample to the $0<z<1$ range. The total number of LGRB host galaxies with measured metallicity is 42 and they are listed in Table~\ref{tab2}.
As shown in Figure~\ref{fig13}, only nine LGRB host galaxies ($21\%$) have super-solar metallicities, while most are well below the solar value.

\begin{figure}
\begin{center}
\includegraphics[angle=0,width=0.8\textwidth]{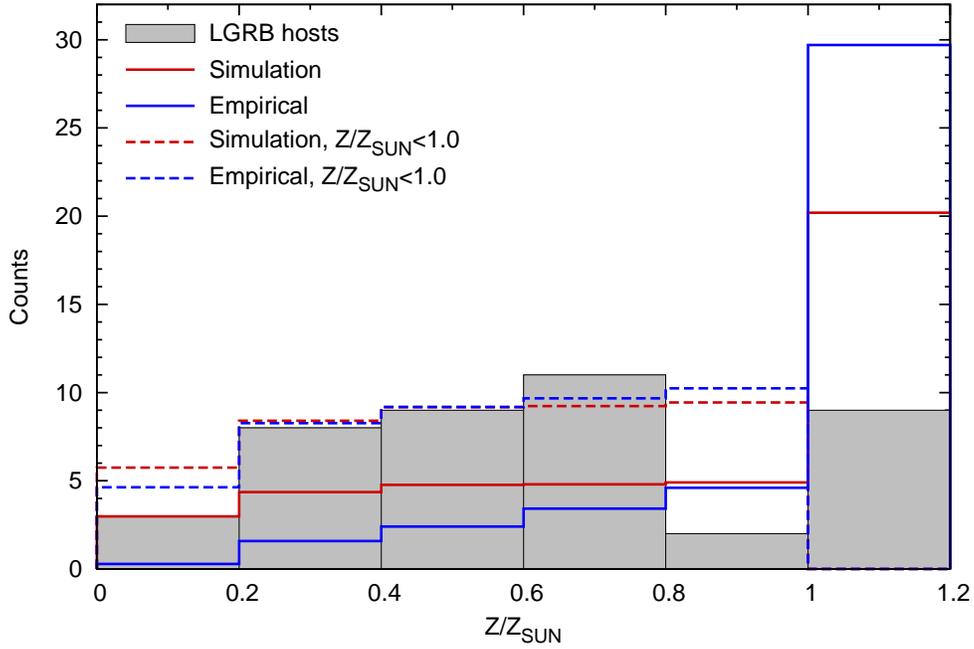}
\end{center}
\caption{LGRB host galaxy metallicity distribution between $0<z\leqslant1$.
The red and blue histograms are the expectations from the Illustris simulation and the empirical model, respectively, assuming LGRBs trace all star formation equally,
while the red and blue dashed ones represent the distributions with a sharp metallicity cutoff of
$Z_{\mathrm{th}}=\mathrm{Z_{\odot}}$.\label{fig14}}
\end{figure}

In Figure~\ref{fig14}, we compare the LGRB host metallicity distribution of our sample from $0<z<1$ with the predictions considering the contribution of the global CSFR, regardless of any bias, in the same redshift range.
The expectations are normalized to the number of the observed hosts.
Note that for the sake of simplicity, all hosts with super-solar metallicities ($Z>Z_{\odot}$) are plotted within the metallicity bin of $1<Z/Z_{\odot}<1.2$.
As can be seen, for the Illustris simulation, the fraction of star formation above the solar metallicity is around $50\%$.
The empirical model shows a similar result as well.
Therefore, if LGRBs trace the star formation unbiasedly all the time,
a significant fraction of their hosts should have super-solar metallicities, which is clearly inconsistent with the observations.
These results confirm previous studies that, at least at lower redshifts, LGRB hosts prefer to occur in
environments below the solar metallicity \citep[e.g.][]{2013ApJ...774..119G,2015A&A...581A.102V,2015ApJ...808...73S,2016ApJ...817....8P,2017MNRAS.469.4921B}.
After considering a sharp metallicity threshold of $Z_{\mathrm{th}}=Z_{\odot}$, we recalculated the CSFRs as a function of stellar metallicity, as shown in Figure~\ref{fig14}. The peaks of the predicted distribution shift from $>Z_{\odot}$ to $\sim0.8\,Z_{\odot}$,
which is much more consistent with the observational data. These results are roughly consistent with that found in the previous section by comparing the cumulative redshift distributions of LGRBs. Here for simplicity in these models, we adopted a sharp cutoff to represent the low-metallicity preference of LGRBs, but we note that LGRBs do occur in
high-metallicity environments, which could be the results of the internal metallicity dispersion in
galaxies \citep{2011MNRAS.417..567N,2017MNRAS.469.4921B}, or may imply the existence of two channels of LGRB progenitors \citep{2015ApJ...802..103T}.

\section{Conclusions and discussion}

In this paper, we have revisited the relationship between the LGRB
rate and CSFR. For this purpose, we collected a large sample of 371
Swift LGRBs with known redshifts and prompt emission properties,
which include the $T_{90}$ duration, fluence and peak photon flux.
A fluence cut of $S_{15-150\,\mathrm{keV}}\geqslant10^{-6}\,\mathrm{erg\, cm^{-2}}$
is used to reduce the selection effect of the sensitivity of the detectors.
We pay special attention to the redshift evolution of the intrinsic emission properties of LGRBs,
such as the isotropic luminosity and the isotropic energy.
Based on our relatively large sample, we could investigate the cumulative luminosity
distributions of LGRBs with $L_{\mathrm{iso}}\gtrsim10^{51}\,\mathrm{erg\, s^{-1}}$
in various redshift bins, finding that the distributions of these high-luminosity bursts
show little evolution with redshift over $1<z<4$.
Therefore, for reproducing the LGRB rate history in this paper,
we choose to ignore the redshift evolution of LGRB luminosity function.
Although given the influences of many unknown observational biases such as the GRB
jet opening angle \citep[e.g.][]{2012ApJ...745..168L},
a more thorough investigation on the redshift evolution of
the luminosity distribution is still needed in the future.

Using the CSFR from the Illustris simulation,
which is a large-scale cosmological hydrodynamical simulation,
we have calculated the expected cumulative
redshift distribution of LGRBs under the assumption of
the existence of host galaxies with different metallicity thresholds.
The advantage of this kind of galaxy formation simulation is to allow us to consider the star
formation and metallicity evolution of each individual galaxy in a self-consistent way.
After comparing with our Swift sample,
we find that all the predictions with the moderate threshold
between $0.3\,Z_{\odot}\leqslant Z_{\mathrm{th}}\leqslant1.0\,Z_{\odot}$
can fit the data well at the redshift range between $0<z<3$.
Models with no metallicity threshold or with an extreme threshold of
$Z_{\mathrm{th}}\leqslant0.2\,Z_{\odot}$ are disfavored.
This is in rough agreement with previous studies arguing that at low redshifts,
LGRBs have a tendency to occur in galaxies with a metallicity below the solar value
\citep[e.g.][]{2008AJ....135.1136M,2012ApJ...744...95R,2014ApJS..213...15W,2015A&A...581A.102V,2017MNRAS.469.4921B}.
At higher redshifts over $3<z<5$, all metallicity thresholds can pass the K-S tests,
implying that the observational data are consistent with no metallicity cutoff.
Since most stars are born in environments of $Z\lesssim 0.6\,Z_{\odot}$ at $z>3$,
the low-metallicity preference would disappear above this redshift given a modest
threshold of $Z_{\mathrm{th}}$.
This may suggest that LGRBs trace all star formation directly at $z>3$.
The predictions from the empirical model also show similar results.
Nevertheless, we caution that it may not be sufficient to claim a strict
constraint on the metallicity threshold only through
the comparison of cumulative LGRB redshift distributions,
since models with a large range in metallicity thresholds exhibit similar
consistency with the data.

To further test the low-metallicity preference of LGRBs, we have also calculated the metallicity
distributions of star forming galaxies from different models and compared them with
the observed LGRB host galaxies.
Based on the results from the Illustris simulation and the empirical model,
we confirm that at $z<1$ a significant fraction of star forming galaxies are in super-solar environments ($\gtrsim50\%$),
which is clearly inconsistent with the metallicity distribution of the observed LGRB host galaxies.
This result implies that the production of LGRBs in these high-metallicity galaxies is somehow suppressed.
We also find that, by considering a metallicity threshold with $Z_{\mathrm{th}}\lesssim Z_{\odot}$,
this discrepancy could be partially alleviated.

Combining these results with the findings from the comparison of the redshift distribution of LGRBs,
we conclude that at low redshifts LGRBs are more likely to occur in galaxies with metallicity below the solar value, consistent with the results of recent work \citep[e.g.][]{2016ApJ...817....8P}, which is higher than some of the previous theoretical works.
While at $z\gtrsim3$, given that a significant fraction of stars are born in sub-solar environments, LGRBs are good enough to
be used as an unbiased tracer of the CSFR, at least to a first approximation.

Furthermore, more observations on LGRB host galaxies, especially at high redshifts,
are desirable to constrain the theoretical models. Meanwhile, the
detection of faint galaxies will also be needed to provide a detailed
picture of the relationship between the LGRB and star formation. Given
the difficulty of observing these objects at high redshifts from current
detectors, next-generation observational facilities, such as the James Webb Space Telescope
are anticipated to provide great insight into the nature of this relationship.

\acknowledgements

We thank the anonymous referee for her/his useful comments and suggestions that
were helpful for significantly improving the manuscript.
This work is partially supported by the National Natural Science Foundation
of China (grant Nos. 11503004, 11873056, 11390372, 11690024, 11991052, 11733001, 11725312, U1431228, 11421303, 11403006, and U1831119),
and the Science and Technology Program of Guangzhou (201707010401).

\bibliographystyle{apj}
\bibliography{refgrb}

\clearpage

\clearpage

\begin{deluxetable}{lcccccccc}
\tabletypesize{\scriptsize}
\tablewidth{0pt}
\tablecaption{The List of 371 Swift LGRBs\label{tab1}}
\tablehead{
\colhead{GRB} & \colhead{$z$} & \colhead{$T_{90}$} & \colhead{$S$} &
\colhead{$P$} & \colhead{$\log(E_{\mathrm{iso}})$} & \colhead{$\log(L_{\mathrm{iso}})$} & \colhead{Refs.}\\
 & & \colhead{(s)} & \colhead{($10^{-7}\,\mathrm{erg\, cm^{-2}}$)} &
 \colhead{($\mathrm{ph\, s^{-1}\, cm^{-2}}$)} & \colhead{($\mathrm{erg}$)} & \colhead{($\mathrm{erg\, s^{-1}}$)} & }
\startdata
GRB171222A & 2.409 & 173.87  & 18.80  & 0.67  & 52.39  & 50.68  & 1\\
GRB171205A & 0.037 & 190.47  & 35.30  & 0.95  & 49.25  & 46.99  & 1\\
GRB171020A & 1.87 & 41.86  & 11.60  & 0.73  & 52.02  & 50.86  & 1\\
GRB170903A & 0.886 & 27.72  & 23.60  & 3.91  & 51.79  & 50.62  & 1\\
GRB170705A & 2.01 & 223.20  & 92.20  & 13.56  & 52.97  & 51.10  & 1\\
GRB170607A & 0.557 & 320.00  & 76.30  & 2.71  & 51.93  & 49.62  & 1\\
GRB170604A & 1.329 & 26.53  & 50.90  & 4.25  & 52.43  & 51.37  & 1\\
GRB170531B & 2.366 & 170.30  & 19.20  & 0.75  & 52.39  & 50.68  & 1\\
GRB170519A & 0.818 & 220.25  & 11.20  & 0.68  & 51.41  & 49.32  & 1\\
GRB170405A & 3.51 & 165.31  & 38.80  & \nodata  & 52.92  & 51.35  & 1\\
GRB170202A & 3.645 & 37.76  & 33.40  & 4.75  & 52.87  & 51.96  & 1\\
GRB170113A & 1.968 & 20.30  & 6.63  & 1.07  & 51.81  & 50.97  & 1\\
GRB161219B & 0.1475 & 6.93  & 15.40  & 5.30  & 50.10  & 49.32  & 1\\
GRB161129A & 0.645 & 35.54  & 35.90  & 3.31  & 51.72  & 50.39  & 4\\
GRB161117A & 1.549 & 125.70  & 202.00  & 6.81  & 53.13  & 51.44  & 1\\
GRB161108A & 1.159 & 115.84  & 11.00  & 0.60  & 51.66  & 49.93  & 1\\
GRB161017A & 2.013 & 217.05  & 53.70  & 2.81  & 52.73  & 50.88  & 1\\
GRB161014A & 2.823 & 23.00  & 22.10  & 2.68  & 52.55  & 51.77  & 1\\
GRB160804A & 0.736 & 152.74  & 109.00  & 2.84  & 52.31  & 50.37  & 1\\
GRB160425A & 0.555 & 304.60  & 20.10  & 2.83  & 51.35  & 49.05  & 1\\
GRB160327A & 4.99 & 33.74  & 13.60  & 1.70  & 52.64  & 51.89  & 1\\
GRB160314A & 0.726 & 8.73  & 2.75  & 0.91  & 50.70  & 50.00  & 1\\
GRB160228A & 1.64 & 98.91  & 19.30  & 1.17  & 52.15  & 50.58  & 3\\
GRB160227A & 2.38 & 316.35  & 31.20  & 0.61  & 52.60  & 50.63  & 1\\
GRB160203A & 3.52 & 17.44  & 9.93  & 1.29  & 52.33  & 51.74  & 1\\
GRB160131A & 0.972 & 327.75  & 201.00  & 0.30  & 52.79  & 50.57  & 1\\
GRB160121A & 1.96 & 10.50  & 6.13  & 1.18  & 51.77  & 51.22  & 1\\
GRB160117B & 0.87 & 11.54  & 3.38  & 1.73  & 50.93  & 50.14  & 1\\
GRB151215A & 2.59 & 17.85  & 3.08  & 1.59  & 51.65  & 50.95  & 1\\
GRB151031A & 1.167 & 5.00  & 3.37  & 1.72  & 51.15  & 50.79  & 1\\
GRB151029A & 1.423 & 8.95  & 4.37  & 1.73  & 51.41  & 50.84  & 1\\
GRB151027B & 4.063 & 80.00  & 14.70  & 0.25  & 52.57  & 51.38  & 1\\
GRB151027A & 0.81 & 129.58  & 72.90  & 6.85  & 52.21  & 50.36  & 1\\
GRB151021A & 2.33 & 110.06  & 274.00  & 9.74  & 53.53  & 52.01  & 1\\
GRB150915A & 1.968 & 160.00  & 7.38  & 0.18  & 51.86  & 50.12  & 1\\
GRB150910A & 1.36 & 112.27  & 47.70  & 1.07  & 52.42  & 50.74  & 1\\
GRB150821A & 0.755 & 168.94  & 8.44  & \nodata  & 51.22  & 49.24  & 1\\
GRB150818A & 0.282 & 143.06  & 42.60  & 2.35  & 51.10  & 49.05  & 1\\
GRB150727A & 0.313 & 87.96  & 36.10  & 1.04  & 51.12  & 49.29  & 1\\
GRB150413A & 3.139 & 243.60  & 42.70  & 1.67  & 52.90  & 51.13  & 1\\
GRB150403A & 2.06 & 37.30  & 171.00  & 0.98  & 53.25  & 52.16  & 1\\
GRB150323A & 0.593 & 149.73  & 56.00  & 4.61  & 51.85  & 49.87  & 1\\
GRB150314A & 1.758 & 14.78  & 225.00  & 37.99  & 53.27  & 52.54  & 1\\
GRB150301B & 1.5169 & 17.14  & 19.80  & 3.01  & 52.11  & 51.28  & 1\\
GRB150206A & 2.087 & 75.00  & 148.00  & 10.06  & 53.20  & 51.81  & 1\\
GRB150120B & 3.5 & 24.34  & 5.34  & 1.22  & 52.06  & 51.32  & 3\\
GRB141225A & 0.915 & 86.15  & 28.10  & 1.35  & 51.89  & 50.24  & 1\\
GRB141221A & 1.452 & 36.82  & 21.10  & 3.06  & 52.11  & 50.93  & 1\\
GRB141220A & 1.32 & 7.23  & 25.60  & 0.26  & 52.12  & 51.63  & 1\\
GRB141121A & 1.47 & 481.00  & 43.20  & 0.88  & 52.43  & 50.14  & 1\\
GRB141109A & 2.993 & 200.19  & 66.60  & 2.41  & 53.06  & 51.36  & 1\\
GRB141026A & 3.35 & 139.48  & 12.90  & 0.36  & 52.41  & 50.91  & 1\\
GRB141004A & 0.573 & 3.92  & 6.68  & 6.13  & 50.89  & 50.50  & 1\\
GRB140907A & 1.21 & 80.00  & 45.10  & 1.57  & 52.31  & 50.75  & 1\\
GRB140710A & 0.558 & 3.00  & 2.29  & 1.48  & 50.41  & 50.12  & 1\\
GRB140703A & 3.14 & 68.64  & 38.80  & 2.77  & 52.86  & 51.64  & 1\\
GRB140629A & 2.275 & 38.27  & 22.90  & 4.32  & 52.44  & 51.37  & 1\\
GRB140614A & 4.233 & 77.39  & 10.30  & 0.32  & 52.44  & 51.27  & 1\\
GRB140518A & 4.707 & 60.52  & 10.60  & 0.97  & 52.51  & 51.48  & 1\\
GRB140515A & 6.33 & 23.42  & 6.19  & 0.91  & 52.41  & 51.91  & 1\\
GRB140512A & 0.725 & 154.11  & 130.00  & 6.05  & 52.37  & 50.42  & 1\\
GRB140509A & 2.4 & 23.22  & 12.30  & 1.55  & 52.20  & 51.37  & 1\\
GRB140506A & 0.889 & 111.10  & 26.60  & 10.64  & 51.85  & 50.08  & 1\\
GRB140430A & 1.6 & 173.59  & 11.30  & 2.50  & 51.90  & 50.08  & 1\\
GRB140428A & 4.7 & 17.42  & 3.40  & 0.64  & 52.01  & 51.53  & 1\\
GRB140423A & 3.26 & 134.14  & 93.70  & 2.15  & 53.26  & 51.76  & 1\\
GRB140419A & 3.956 & 80.08  & 157.00  & 4.50  & 53.59  & 52.38  & 1\\
GRB140331A & 4.65 & 209.66  & 7.13  & 0.22  & 52.33  & 50.76  & 1\\
GRB140318A & 1.02 & 7.60  & 2.85  & 0.54  & 50.98  & 50.41  & 1\\
GRB140311A & 4.954 & 70.48  & 21.00  & 1.28  & 52.83  & 51.75  & 1\\
GRB140304A & 5.283 & 14.78  & 11.00  & 1.66  & 52.58  & 52.21  & 1\\
GRB140301A & 1.416 & 27.80  & 4.55  & 0.81  & 51.42  & 50.36  & 1\\
GRB140213A & 1.2076 & 59.93  & 120.00  & 23.55  & 52.73  & 51.30  & 1\\
GRB140206A & 2.73 & 94.19  & 164.00  & 19.95  & 53.40  & 52.00  & 1\\
GRB140114A & 3 & 139.95  & 32.00  & 0.95  & 52.75  & 51.20  & 1\\
GRB131227A & 5.3 & 17.99  & 8.41  & 1.15  & 52.46  & 52.01  & 1\\
GRB131117A & 4.042 & 10.88  & 2.98  & 0.66  & 51.88  & 51.54  & 1\\
GRB131103A & 0.5955 & 15.21  & 8.12  & 1.56  & 51.01  & 50.03  & 1\\
GRB131030A & 1.293 & 39.42  & 291.00  & 28.09  & 53.17  & 51.93  & 1\\
GRB130907A & 1.238 & 364.37  & 1360.00  & 1.27  & 53.80  & 51.59  & 1\\
GRB130831A & 0.4791 & 30.19  & 65.00  & 13.56  & 51.73  & 50.42  & 1\\
GRB130701A & 1.155 & 4.38  & 43.80  & 17.10  & 52.26  & 51.95  & 1\\
GRB130612A & 2.006 & 4.00  & 2.81  & 1.62  & 51.45  & 51.33  & 1\\
GRB130610A & 2.092 & 47.72  & 25.30  & 1.72  & 52.43  & 51.24  & 1\\
GRB130606A & 5.9134 & 276.66  & 27.60  & 2.51  & 53.03  & 51.43  & 1\\
GRB130604A & 1.06 & 76.28  & 15.40  & 0.77  & 51.74  & 50.17  & 1\\
GRB130528A & 1.25 & 640.00  & 56.80  & 0.25  & 52.43  & 49.98  & 1\\
GRB130514A & 3.6 & 214.19  & 90.00  & 2.93  & 53.30  & 51.63  & 1\\
GRB130511A & 1.3033 & 2.74  & 1.74  & 1.26  & 50.95  & 50.87  & 1\\
GRB130505A & 2.27 & 89.34  & 34.70  & \nodata  & 52.62  & 51.18  & 1\\
GRB130427B & 2.78 & 25.90  & 14.50  & 3.15  & 52.36  & 51.52  & 1\\
GRB130427A & 0.3399 & 244.33  & 3730.00  & 358.63  & 53.20  & 50.94  & 1\\
GRB130420A & 1.297 & 121.14  & 75.90  & 3.33  & 52.58  & 50.86  & 1\\
GRB130418A & 1.218 & 274.92  & 17.70  & 0.67  & 51.91  & 49.81  & 1\\
GRB130408A & 3.758 & 4.24  & 16.70  & 5.12  & 52.59  & 52.64  & 1\\
GRB130215A & 0.597 & 66.22  & 54.00  & 2.58  & 51.83  & 50.22  & 1\\
GRB130131B & 2.539 & 4.30  & 3.38  & 0.96  & 51.67  & 51.59  & 1\\
GRB121229A & 2.707 & 111.46  & 8.40  & 0.43  & 52.11  & 50.63  & 1\\
GRB121217A & 3.1 & 778.09  & 61.70  & 1.75  & 53.05  & 50.77  & 1\\
GRB121211A & 1.023 & 182.70  & 12.70  & 1.02  & 51.63  & 49.68  & 1\\
GRB121209A & 2.1 & 42.92  & 28.70  & 3.38  & 52.49  & 51.35  & 1\\
GRB121201A & 3.385 & 38.00  & 7.20  & 0.80  & 52.17  & 51.23  & 1\\
GRB121128A & 2.2 & 23.43  & 58.30  & 12.46  & 52.82  & 51.96  & 1\\
GRB121024A & 2.298 & 67.97  & 11.00  & 1.37  & 52.13  & 50.81  & 1\\
GRB120923A & 7.84 & 26.08  & 3.89  & 0.63  & 52.30  & 51.83  & 1\\
GRB120922A & 3.1 & 168.22  & 55.20  & 1.69  & 53.00  & 51.39  & 1\\
GRB120909A & 3.93 & 220.60  & 75.00  & 1.81  & 53.26  & 51.61  & 1\\
GRB120907A & 0.97 & 6.08  & 5.88  & 2.86  & 51.26  & 50.77  & 1\\
GRB120815A & 2.3586 & 7.23  & 4.92  & 2.23  & 51.79  & 51.46  & 1\\
GRB120811C & 2.671 & 24.34  & 28.40  & 3.96  & 52.63  & 51.81  & 1\\
GRB120805A & 3.1 & 48.00  & 8.41  & 0.17  & 52.19  & 51.12  & 1\\
GRB120802A & 3.796 & 50.29  & 16.40  & 3.03  & 52.59  & 51.57  & 1\\
GRB120729A & 0.8 & 93.93  & 25.10  & 2.81  & 51.74  & 50.02  & 1\\
GRB120724A & 1.48 & 77.92  & 8.00  & 0.72  & 51.70  & 50.20  & 1\\
GRB120722A & 0.9586 & 36.32  & 13.30  & 1.00  & 51.60  & 50.34  & 1\\
GRB120714B & 0.3984 & 157.31  & 12.50  & 0.35  & 50.86  & 48.81  & 1\\
GRB120712A & 4.1745 & 14.81  & 18.40  & 2.30  & 52.69  & 52.23  & 1\\
GRB120624B & 2.1974 & 179.66  & 296.00  & 5.39  & 53.53  & 51.78  & 1\\
GRB120521C & 5.93 & 27.07  & 12.00  & 1.92  & 52.67  & 52.08  & 1\\
GRB120422A & 0.28253 & 60.35  & 3.22  & 0.57  & 49.98  & 48.31  & 1\\
GRB120404A & 2.876 & 38.72  & 15.80  & 1.16  & 52.42  & 51.42  & 1\\
GRB120401A & 4.5 & 130.27  & 9.21  & 0.38  & 52.42  & 51.05  & 1\\
GRB120327A & 2.8145 & 63.53  & 35.70  & 3.88  & 52.76  & 51.54  & 1\\
GRB120326A & 1.798 & 69.48  & 25.20  & 4.74  & 52.33  & 50.94  & 1\\
GRB120224A & 1.1 & 7.00  & 2.16  & 0.85  & 50.92  & 50.40  & 1\\
GRB120211A & 2.4 & 64.32  & 8.62  & 0.46  & 52.05  & 50.77  & 1\\
GRB120119A & 1.728 & 68.00  & 170.00  & 9.69  & 53.13  & 51.74  & 1\\
GRB120118B & 2.943 & 20.30  & 17.50  & 2.16  & 52.48  & 51.76  & 1\\
GRB111229A & 1.3805 & 25.37  & 3.36  & 0.97  & 51.28  & 50.25  & 1\\
GRB111228A & 0.71627 & 101.24  & 82.70  & 12.23  & 52.17  & 50.40  & 1\\
GRB111225A & 0.297 & 105.73  & 12.60  & 0.65  & 50.61  & 48.70  & 1\\
GRB111129A & 1.0796 & 8.48  & 1.82  & 0.92  & 50.83  & 50.22  & 1\\
GRB111123A & 3.1516 & 290.00  & 71.00  & 0.89  & 53.12  & 51.28  & 1\\
GRB111107A & 2.893 & 31.07  & 9.12  & 1.24  & 52.18  & 51.28  & 1\\
GRB111008A & 4.99005 & 62.85  & 52.00  & 6.42  & 53.22  & 52.20  & 1\\
GRB111005A & 0.01326 & 23.21  & 5.75  & 1.13  & 47.56  & 46.20  & 1\\
GRB110818A & 3.36 & 102.84  & 39.90  & 1.57  & 52.91  & 51.53  & 1\\
GRB110808A & 1.348 & 40.70  & 3.69  & 0.58  & 51.30  & 50.06  & 1\\
GRB110801A & 1.858 & 385.26  & 45.60  & 1.01  & 52.61  & 50.48  & 1\\
GRB110731A & 2.83 & 40.94  & 59.40  & 10.67  & 52.98  & 51.95  & 1\\
GRB110726A & 1.036 & 5.16  & 2.45  & 1.01  & 50.93  & 50.52  & 1\\
GRB110715A & 0.8224 & 13.00  & 120.00  & 52.42  & 52.44  & 51.59  & 1\\
GRB110503A & 1.613 & 58.70  & 113.00  & 29.63  & 52.91  & 51.56  & 1\\
GRB110422A & 1.77 & 25.78  & 383.00  & 28.88  & 53.50  & 52.53  & 1\\
GRB110213A & 1.4607 & 48.00  & 59.70  & 1.56  & 52.56  & 51.27  & 1\\
GRB110205A & 2.22 & 249.42  & 158.00  & 3.12  & 53.26  & 51.37  & 1\\
GRB110128A & 2.339 & 14.15  & 5.78  & 0.76  & 51.86  & 51.23  & 1\\
GRB110106B & 0.618 & 43.42  & 20.50  & 2.06  & 51.44  & 50.01  & 1\\
GRB101219B & 0.55185 & 41.86  & 25.90  & 1.67  & 51.45  & 50.02  & 1\\
GRB101213A & 0.414 & 131.12  & 49.90  & 2.27  & 51.50  & 49.53  & 1\\
GRB100906A & 1.727 & 114.63  & 119.00  & 10.02  & 52.98  & 51.35  & 1\\
GRB100902A & 4.5 & 428.83  & 32.20  & 0.96  & 52.97  & 51.07  & 4\\
GRB100901A & 1.4084 & 436.66  & 19.50  & 0.79  & 52.05  & 49.79  & 1\\
GRB100814A & 1.44 & 177.26  & 88.50  & 2.47  & 52.73  & 50.86  & 1\\
GRB100728B & 2.106 & 12.08  & 16.70  & 3.47  & 52.25  & 51.66  & 1\\
GRB100728A & 1.567 & 193.38  & 373.00  & 5.00  & 53.41  & 51.53  & 1\\
GRB100704A & 3.6 & 196.88  & 58.50  & 4.30  & 53.11  & 51.48  & 4\\
GRB100621A & 0.542 & 63.55  & 206.00  & 12.61  & 52.34  & 50.72  & 1\\
GRB100615A & 1.398 & 38.82  & 49.20  & 5.56  & 52.45  & 51.24  & 1\\
GRB100513A & 4.772 & 83.50  & 14.30  & 0.57  & 52.64  & 51.48  & 1\\
GRB100508A & 0.5201 & 49.26  & 7.50  & 0.47  & 50.86  & 49.35  & 1\\
GRB100425A & 1.755 & 38.97  & 4.64  & 1.37  & 51.58  & 50.43  & 1\\
GRB100424A & 2.465 & 104.00  & 15.00  & 0.33  & 52.30  & 50.83  & 1\\
GRB100418A & 0.6239 & 7.93  & 3.46  & 1.04  & 50.68  & 49.99  & 1\\
GRB100413A & 3.9 & 192.64  & 61.80  & 0.64  & 53.18  & 51.58  & 4\\
GRB100316B & 1.18 & 3.84  & 1.98  & 1.34  & 50.93  & 50.69  & 1\\
GRB100316A & 3.155 & 6.75  & 8.22  & 2.26  & 52.19  & 51.98  & 1\\
GRB100302A & 4.813 & 17.95  & 3.16  & 0.47  & 51.99  & 51.50  & 1\\
GRB100219A & 4.66723 & 27.57  & 4.65  & 0.40  & 52.14  & 51.46  & 1\\
GRB100213B & 0.604 & 91.86  & 12.20  & 0.80  & 51.20  & 49.44  & 1\\
GRB091208B & 1.0633 & 14.80  & 32.00  & 14.62  & 52.06  & 51.21  & 1\\
GRB091127 & 0.49044 & 6.96  & 85.80  & 45.31  & 51.87  & 51.20  & 1\\
GRB091109A & 3.076 & 48.03  & 16.40  & 1.23  & 52.47  & 51.40  & 1\\
GRB091029 & 2.752 & 39.18  & 24.10  & 1.72  & 52.58  & 51.56  & 1\\
GRB091024 & 1.092 & 112.28  & 59.80  & 2.06  & 52.35  & 50.62  & 1\\
GRB091020 & 1.71 & 38.92  & 38.00  & 4.18  & 52.48  & 51.32  & 1\\
GRB091018 & 0.971 & 4.37  & 15.20  & 10.38  & 51.67  & 51.33  & 1\\
GRB090927 & 1.37 & 2.16  & 1.98  & 1.90  & 51.04  & 51.08  & 1\\
GRB090926B & 1.24 & 99.28  & 71.20  & 3.16  & 52.52  & 50.88  & 1\\
GRB090814A & 0.696 & 78.06  & 12.50  & 0.54  & 51.32  & 49.66  & 1\\
GRB090812 & 2.452 & 74.50  & 57.30  & 3.55  & 52.88  & 51.55  & 1\\
GRB090809A & 2.737 & 8.87  & 3.65  & 1.08  & 51.75  & 51.38  & 1\\
GRB090726 & 2.71 & 56.68  & 7.86  & 0.79  & 52.08  & 50.90  & 1\\
GRB090715B & 3 & 266.40  & 56.50  & 3.75  & 52.99  & 51.17  & 1\\
GRB090709A & 1.8 & 88.74  & 253.00  & 7.79  & 53.33  & 51.83  & 1\\
GRB090618 & 0.54 & 113.34  & 1090.00  & 1.73  & 53.06  & 51.19  & 1\\
GRB090530 & 1.266 & 40.46  & 10.90  & 3.19  & 51.72  & 50.47  & 1\\
GRB090529A & 2.625 & 70.44  & 10.40  & 0.69  & 52.18  & 50.89  & 1\\
GRB090519 & 3.85 & 58.04  & 11.80  & 0.57  & 52.45  & 51.37  & 1\\
GRB090516A & 4.109 & 181.01  & 92.10  & 2.31  & 53.38  & 51.83  & 1\\
GRB090429B & 9.38 & 5.58  & 3.29  & 1.63  & 52.30  & 52.57  & 1\\
GRB090424 & 0.544 & 49.46  & 218.00  & 69.92  & 52.36  & 50.86  & 1\\
GRB090423 & 8.26 & 10.30  & 6.24  & 1.74  & 52.53  & 52.48  & 1\\
GRB090418A & 1.608 & 56.30  & 46.60  & 1.90  & 52.52  & 51.19  & 1\\
GRB090407 & 1.4485 & 315.47  & 11.40  & 0.66  & 51.84  & 49.73  & 1\\
GRB090404 & 2.87 & 82.02  & 31.20  & 1.97  & 52.71  & 51.39  & 1\\
GRB090313 & 3.3736 & 83.04  & 15.10  & 0.75  & 52.49  & 51.21  & 1\\
GRB090205 & 4.6497 & 8.81  & 2.03  & 0.48  & 51.78  & 51.59  & 1\\
GRB090201 & 2.1 & 74.26  & 289.00  & 14.57  & 53.49  & 52.11  & 1\\
GRB090113 & 1.7493 & 9.10  & 7.75  & 2.47  & 51.80  & 51.28  & 1\\
GRB090102 & 1.547 & 28.32  & 70.60  & 0.21  & 52.68  & 51.63  & 1\\
GRB081230 & 2.03 & 60.69  & 8.59  & 0.71  & 51.94  & 50.64  & 1\\
GRB081228 & 3.44 & 3.00  & 0.93  & 0.62  & 51.29  & 51.46  & 1\\
GRB081222 & 2.77 & 33.00  & 52.20  & 7.46  & 52.91  & 51.97  & 1\\
GRB081221 & 2.26 & 33.91  & 189.00  & 17.95  & 53.35  & 52.33  & 1\\
GRB081210 & 2.0631 & 145.91  & 18.50  & 2.45  & 52.29  & 50.61  & 2\\
GRB081203A & 2.05 & 223.00  & 78.30  & 2.82  & 52.91  & 51.04  & 1\\
GRB081121 & 2.512 & 17.52  & 50.80  & 4.15  & 52.85  & 52.15  & 1\\
GRB081118 & 2.58 & 53.40  & 11.80  & 0.60  & 52.23  & 51.05  & 1\\
GRB081109A & 0.9787 & 221.49  & 40.00  & 1.41  & 52.10  & 50.05  & 1\\
GRB081029 & 3.8479 & 275.10  & 21.40  & 0.45  & 52.71  & 50.95  & 1\\
GRB081028A & 3.038 & 284.42  & 39.50  & 0.57  & 52.85  & 51.00  & 1\\
GRB081008 & 1.9683 & 187.82  & 42.50  & 1.29  & 52.62  & 50.82  & 1\\
GRB081007 & 0.5295 & 9.73  & 7.61  & 2.87  & 50.89  & 50.08  & 1\\
GRB080928 & 1.6919 & 233.66  & 24.30  & 2.08  & 52.27  & 50.34  & 1\\
GRB080916A & 0.6887 & 61.35  & 42.20  & 2.62  & 51.84  & 50.28  & 1\\
GRB080913 & 6.733 & 7.46  & 5.72  & 1.37  & 52.40  & 52.42  & 1\\
GRB080906 & 2.13 & 148.21  & 36.50  & 1.02  & 52.60  & 50.93  & 1\\
GRB080905B & 2.3739 & 120.94  & 18.10  & 1.62  & 52.36  & 50.81  & 1\\
GRB080810 & 3.3604 & 107.67  & 46.40  & 1.98  & 52.97  & 51.58  & 1\\
GRB080805 & 1.5042 & 106.62  & 26.50  & 1.04  & 52.23  & 50.60  & 1\\
GRB080804 & 2.2045 & 37.87  & 37.90  & 3.06  & 52.64  & 51.57  & 1\\
GRB080721 & 2.5914 & 129.70  & 155.00  & 0.71  & 53.35  & 51.79  & 1\\
GRB080710 & 0.8454 & 142.99  & 15.30  & 0.92  & 51.57  & 49.68  & 1\\
GRB080707 & 1.2322 & 30.16  & 6.26  & 1.06  & 51.46  & 50.33  & 1\\
GRB080607 & 3.0368 & 78.97  & 247.00  & 23.30  & 53.64  & 52.35  & 1\\
GRB080605 & 1.6403 & 18.06  & 134.00  & 0.94  & 53.00  & 52.16  & 1\\
GRB080604 & 1.4171 & 77.61  & 7.89  & 0.38  & 51.66  & 50.16  & 1\\
GRB080603B & 2.6892 & 59.12  & 24.60  & 3.53  & 52.57  & 51.37  & 1\\
GRB080602 & 1.8204 & 74.29  & 32.60  & 2.73  & 52.45  & 51.03  & 1\\
GRB080520 & 1.5457 & 3.32  & 0.59  & 0.49  & 50.60  & 50.48  & 1\\
GRB080517 & 0.089 & 64.51  & 5.87  & 0.60  & 49.24  & 47.47  & 1\\
GRB080516 & 3.2 & 5.76  & 2.67  & 1.70  & 51.71  & 51.57  & 1\\
GRB080515 & 2.47 & 20.86  & 25.80  & 3.42  & 52.54  & 51.76  & 1\\
GRB080430 & 0.767 & 13.87  & 11.70  & 2.66  & 51.37  & 50.48  & 1\\
GRB080413B & 1.1014 & 8.00  & 33.60  & 18.13  & 52.11  & 51.53  & 1\\
GRB080413A & 2.433 & 46.36  & 35.20  & 5.52  & 52.67  & 51.54  & 1\\
GRB080411 & 1.0301 & 56.33  & 265.00  & 40.61  & 52.96  & 51.51  & 1\\
GRB080330 & 1.5119 & 60.36  & 2.94  & 0.94  & 51.28  & 49.90  & 1\\
GRB080325 & 1.78 & 166.74  & 50.10  & 1.43  & 52.62  & 50.84  & 1\\
GRB080319C & 1.9492 & 29.55  & 35.10  & 5.09  & 52.53  & 51.53  & 1\\
GRB080319B & 0.9382 & 124.86  & 855.00  & 1.24  & 53.39  & 51.59  & 1\\
GRB080319A & 2.0265 & 43.62  & 44.50  & 2.60  & 52.66  & 51.50  & 2\\
GRB080310 & 2.42743 & 363.21  & 23.10  & 1.32  & 52.48  & 50.46  & 1\\
GRB080210 & 2.6419 & 42.26  & 18.00  & 1.59  & 52.42  & 51.36  & 1\\
GRB080207 & 2.0858 & 292.46  & 64.00  & 1.10  & 52.83  & 50.85  & 1\\
GRB080205 & 2.72 & 106.30  & 20.30  & 1.40  & 52.49  & 51.04  & 2\\
GRB080129 & 4.349 & 50.18  & 8.39  & 0.55  & 52.36  & 51.39  & 1\\
GRB071122 & 1.14 & 71.43  & 6.16  & 0.41  & 51.40  & 49.88  & 1\\
GRB071117 & 1.3308 & 6.08  & 24.20  & 0.42  & 52.11  & 51.69  & 1\\
GRB071031 & 2.6918 & 180.64  & 8.98  & 0.54  & 52.13  & 50.44  & 1\\
GRB071025 & 5.2 & 241.30  & 73.20  & 1.71  & 53.39  & 51.80  & 1\\
GRB071021 & 2.452 & 228.72  & 13.60  & 0.63  & 52.26  & 50.44  & 1\\
GRB071020 & 2.1462 & 4.30  & 23.40  & 0.61  & 52.41  & 52.28  & 1\\
GRB071010B & 0.947 & 36.12  & 46.20  & 7.37  & 52.13  & 50.87  & 1\\
GRB071010A & 0.985 & 6.32  & 2.02  & 0.89  & 50.81  & 50.30  & 1\\
GRB071003 & 1.60435 & 148.39  & 83.50  & 6.35  & 52.77  & 51.02  & 1\\
GRB070810A & 2.17 & 9.04  & 6.72  & 1.89  & 51.88  & 51.42  & 1\\
GRB070808 & 1.35 & 58.43  & 12.90  & 1.94  & 51.84  & 50.45  & 2\\
GRB070802 & 2.4541 & 15.80  & 2.79  & 0.45  & 51.57  & 50.91  & 1\\
GRB070721B & 3.6298 & 336.86  & 35.60  & 1.56  & 52.90  & 51.04  & 1\\
GRB070714A & 1.58 & 3.00  & 1.55  & 1.90  & 51.03  & 50.97  & 1\\
GRB070612A & 0.617 & 365.28  & 108.00  & 1.37  & 52.16  & 49.81  & 1\\
GRB070611 & 2.0394 & 13.18  & 4.03  & 0.80  & 51.62  & 50.98  & 1\\
GRB070529 & 2.4996 & 108.90  & 24.80  & 1.46  & 52.53  & 51.04  & 1\\
GRB070521 & 2.0865 & 38.63  & 82.10  & 0.40  & 52.94  & 51.84  & 1\\
GRB070518 & 1.16 & 5.50  & 1.67  & 0.67  & 50.85  & 50.44  & 1\\
GRB070508 & 0.82 & 20.90  & 201.00  & 24.25  & 52.66  & 51.60  & 1\\
GRB070506 & 2.309 & 5.99  & 2.30  & 0.99  & 51.45  & 51.19  & 1\\
GRB070420 & 3.01 & 77.02  & 142.00  & 7.06  & 53.40  & 52.11  & 1\\
GRB070419B & 1.9588 & 238.01  & 74.40  & 1.36  & 52.86  & 50.95  & 1\\
GRB070419A & 0.9705 & 160.00  & 6.67  & 0.06  & 51.31  & 49.40  & 1\\
GRB070411 & 2.9538 & 115.69  & 26.90  & 0.93  & 52.66  & 51.20  & 1\\
GRB070328 & 2.06 & 72.12  & 91.50  & 4.21  & 52.98  & 51.61  & 1\\
GRB070318 & 0.8397 & 130.37  & 26.20  & 1.61  & 51.80  & 49.94  & 1\\
GRB070306 & 1.49594 & 209.24  & 54.60  & 4.06  & 52.54  & 50.62  & 1\\
GRB070224 & 1.99 & 48.00  & 3.70  & 0.20  & 51.56  & 50.36  & 1\\
GRB070223 & 1.6295 & 128.00  & 17.90  & 0.44  & 52.12  & 50.43  & 2\\
GRB070208 & 1.165 & 64.00  & 5.18  & 0.38  & 51.34  & 49.87  & 1\\
GRB070129 & 2.3384 & 459.75  & 30.00  & 0.56  & 52.57  & 50.43  & 1\\
GRB070110 & 2.3521 & 88.43  & 16.30  & 0.61  & 52.31  & 50.89  & 1\\
GRB070103 & 2.6208 & 18.41  & 3.29  & 1.05  & 51.68  & 50.98  & 1\\
GRB061222B & 3.355 & 37.25  & 22.70  & 1.50  & 52.66  & 51.73  & 1\\
GRB061222A & 2.088 & 100.00  & 85.10  & 7.45  & 52.96  & 51.45  & 1\\
GRB061202 & 2.2543 & 94.19  & 34.50  & 2.47  & 52.61  & 51.15  & 2\\
GRB061126 & 1.1588 & 52.62  & 66.50  & 0.46  & 52.45  & 51.06  & 1\\
GRB061121 & 1.3145 & 81.22  & 139.00  & 1.00  & 52.86  & 51.31  & 1\\
GRB061110B & 3.4344 & 135.25  & 13.50  & 0.46  & 52.45  & 50.96  & 1\\
GRB061110A & 0.7578 & 44.51  & 11.30  & 0.47  & 51.35  & 49.95  & 1\\
GRB061021 & 0.3463 & 47.82  & 30.10  & 6.12  & 51.12  & 49.57  & 1\\
GRB061007 & 1.2622 & 75.74  & 450.00  & 0.96  & 53.34  & 51.81  & 1\\
GRB060927 & 5.4636 & 22.42  & 11.20  & 2.68  & 52.60  & 52.06  & 1\\
GRB060926 & 3.2086 & 8.82  & 2.32  & 1.10  & 51.65  & 51.32  & 1\\
GRB060923B & 1.51 & 8.95  & 4.90  & 1.42  & 51.50  & 50.95  & 1\\
GRB060923A & 2.47 & 58.49  & 9.77  & 1.37  & 52.12  & 50.89  & 1\\
GRB060912A & 0.937 & 5.03  & 13.60  & 8.54  & 51.60  & 51.18  & 1\\
GRB060908 & 1.8836 & 19.30  & 28.10  & 2.98  & 52.41  & 51.58  & 1\\
GRB060906 & 3.6856 & 44.59  & 22.90  & 1.98  & 52.72  & 51.74  & 1\\
GRB060904B & 0.7029 & 189.98  & 17.10  & 2.43  & 51.47  & 49.42  & 1\\
GRB060904A & 2.55 & 80.06  & 78.30  & 4.91  & 53.04  & 51.69  & 1\\
GRB060814 & 1.9229 & 145.07  & 148.00  & 7.31  & 53.14  & 51.45  & 1\\
GRB060805A & 2.3633 & 4.93  & 0.73  & 0.31  & 50.97  & 50.80  & 1\\
GRB060729 & 0.5428 & 113.04  & 26.10  & 1.15  & 51.44  & 49.58  & 1\\
GRB060719 & 1.532 & 66.92  & 15.30  & 2.15  & 52.01  & 50.58  & 1\\
GRB060714 & 2.7108 & 116.06  & 28.80  & 1.27  & 52.64  & 51.15  & 1\\
GRB060708 & 1.92 & 10.03  & 4.99  & 1.91  & 51.67  & 51.13  & 1\\
GRB060707 & 3.424 & 66.64  & 15.80  & 1.08  & 52.51  & 51.34  & 1\\
GRB060614 & 0.1254 & 109.10  & 188.00  & 11.49  & 51.04  & 49.05  & 1\\
GRB060607A & 3.0749 & 103.03  & 25.70  & 1.40  & 52.67  & 51.26  & 1\\
GRB060605 & 3.773 & 79.84  & 7.11  & 0.47  & 52.22  & 51.00  & 1\\
GRB060604 & 2.1357 & 96.00  & 3.76  & 0.02  & 51.62  & 50.13  & 1\\
GRB060602A & 0.787 & 74.84  & 16.10  & 0.56  & 51.53  & 49.91  & 1\\
GRB060526 & 3.2213 & 298.04  & 12.90  & 1.67  & 52.39  & 50.54  & 1\\
GRB060522 & 5.11 & 69.12  & 11.30  & 0.54  & 52.57  & 51.52  & 1\\
GRB060512 & 2.1 & 8.40  & 2.31  & 0.89  & 51.39  & 50.96  & 1\\
GRB060510B & 4.941 & 262.94  & 39.80  & 0.57  & 53.10  & 51.46  & 1\\
GRB060502A & 1.5026 & 28.45  & 23.30  & 1.73  & 52.18  & 51.12  & 1\\
GRB060428B & 0.348 & 96.00  & 8.44  & 0.31  & 50.58  & 48.72  & 1\\
GRB060418 & 1.49 & 109.08  & 84.00  & 6.56  & 52.73  & 51.08  & 1\\
GRB060319 & 1.172 & 10.29  & 2.65  & 1.07  & 51.05  & 50.38  & 1\\
GRB060306 & 1.559 & 60.94  & 21.50  & 5.92  & 52.17  & 50.79  & 1\\
GRB060223A & 4.41 & 11.32  & 6.74  & 1.35  & 52.28  & 51.96  & 1\\
GRB060210 & 3.9122 & 288.00  & 76.80  & 1.20  & 53.27  & 51.50  & 1\\
GRB060206 & 4.0559 & 7.55  & 8.42  & 2.78  & 52.33  & 52.16  & 1\\
GRB060204B & 2.3393 & 139.46  & 29.30  & 1.33  & 52.56  & 50.94  & 2\\
GRB060202 & 0.783 & 192.88  & 22.20  & 0.53  & 51.67  & 49.63  & 1\\
GRB060124 & 2.3 & 13.42  & 4.65  & 0.86  & 51.75  & 51.14  & 1\\
GRB060117 & 4.6 & 16.85  & 210.00  & 48.49  & 53.79  & 53.31  & 1\\
GRB060116 & 6.6 & 104.83  & 23.90  & 1.17  & 53.02  & 51.88  & 1\\
GRB060115 & 3.5328 & 139.09  & 17.10  & 0.86  & 52.57  & 51.08  & 1\\
GRB060111A & 2.32 & 13.21  & 12.00  & 1.71  & 52.17  & 51.57  & 1\\
GRB060108 & 2.03 & 14.22  & 3.72  & 0.75  & 51.58  & 50.91  & 1\\
GRB051117B & 0.481 & 9.02  & 1.75  & 0.50  & 50.17  & 49.38  & 1\\
GRB051111 & 1.54948 & 64.00  & 42.70  & 1.83  & 52.46  & 51.06  & 1\\
GRB051109B & 0.08 & 15.70  & 2.65  & 0.57  & 48.80  & 47.64  & 1\\
GRB051109A & 2.346 & 37.20  & 21.70  & 3.88  & 52.43  & 51.39  & 1\\
GRB051016B & 0.9364 & 4.00  & 1.67  & 1.29  & 50.68  & 50.37  & 1\\
GRB051008 & 2.77 & 12.41  & 50.80  & 0.19  & 52.90  & 52.39  & 1\\
GRB051006 & 1.059 & 35.41  & 13.50  & 1.68  & 51.69  & 50.45  & 1\\
GRB051001 & 2.4296 & 190.26  & 17.60  & 0.48  & 52.37  & 50.62  & 1\\
GRB050922C & 2.1995 & 4.55  & 16.20  & 7.09  & 52.27  & 52.12  & 1\\
GRB050922B & 4.5 & 157.02  & 23.70  & 1.09  & 52.83  & 51.38  & 1\\
GRB050915A & 2.5273 & 53.42  & 8.36  & 0.76  & 52.07  & 50.89  & 1\\
GRB050908 & 3.3467 & 18.28  & 4.83  & 0.70  & 51.99  & 51.36  & 1\\
GRB050904 & 6.295 & 181.58  & 52.10  & 0.63  & 53.33  & 51.94  & 1\\
GRB050826 & 0.296 & 29.60  & 4.20  & 0.37  & 50.13  & 48.78  & 1\\
GRB050824 & 0.8278 & 25.01  & 2.78  & 0.48  & 50.81  & 49.67  & 1\\
GRB050822 & 1.434 & 104.29  & 25.10  & 2.23  & 52.18  & 50.54  & 1\\
GRB050820A & 2.6147 & 240.77  & 38.30  & 2.44  & 52.75  & 50.92  & 1\\
GRB050819 & 2.5043 & 37.72  & 3.48  & 0.40  & 51.68  & 50.65  & 1\\
GRB050814 & 5.3 & 142.85  & 19.10  & 0.67  & 52.82  & 51.46  & 1\\
GRB050803 & 3.5 & 88.12  & 21.60  & 0.96  & 52.66  & 51.37  & 1\\
GRB050802 & 1.7102 & 27.46  & 22.10  & 2.66  & 52.24  & 51.23  & 1\\
GRB050801 & 1.38 & 19.57  & 3.09  & 1.47  & 51.24  & 50.32  & 1\\
GRB050730 & 3.9693 & 154.60  & 23.60  & 0.56  & 52.77  & 51.27  & 1\\
GRB050726 & 0.1646 & 46.50  & 19.60  & 1.37  & 50.30  & 48.70  & 1\\
GRB050714B & 2.438 & 49.36  & 6.30  & 0.55  & 51.92  & 50.76  & 1\\
GRB050603 & 2.821 & 21.00  & 82.10  & 0.76  & 53.12  & 52.38  & 1\\
GRB050525A & 0.606 & 8.84  & 151.00  & 43.85  & 52.29  & 51.55  & 1\\
GRB050505 & 4.2748 & 58.85  & 24.90  & 1.84  & 52.83  & 51.78  & 1\\
GRB050502B & 5.2 & 17.72  & 4.72  & 1.39  & 52.20  & 51.75  & 1\\
GRB050416A & 0.6528 & 6.67  & 4.13  & 4.86  & 50.79  & 50.19  & 1\\
GRB050406 & 2.7 & 5.78  & 0.78  & 0.34  & 51.07  & 50.88  & 1\\
GRB050401 & 2.8983 & 32.09  & 80.90  & 12.17  & 53.13  & 52.22  & 1\\
GRB050319 & 3.2425 & 151.58  & 13.30  & 1.46  & 52.41  & 50.86  & 1\\
GRB050318 & 1.4436 & 16.12  & 10.40  & 3.13  & 51.80  & 50.98  & 1\\
GRB050315 & 1.95 & 95.40  & 31.70  & 1.89  & 52.48  & 50.97  & 1\\
GRB050223 & 0.584 & 22.68  & 6.48  & 0.65  & 50.90  & 49.74  & 1\\
GRB050219A & 0.2115 & 23.81  & 40.80  & 3.43  & 50.83  & 49.54  & 1\\
GRB050215B & 2.52 & 11.04  & 2.35  & 0.67  & 51.51  & 51.02  & 1\\
GRB050126 & 1.29 & 48.00  & 8.85  & 0.31  & 51.65  & 50.33  & 1\\
\enddata
\tablecomments{We give the name, redshift $z$, duration $T_{90}$, fluence $S$ in 15-150~keV,
peak photon flux $P$, isotropic
energy $E_{\mathrm{iso}}$, and luminosity $L_{\mathrm{iso}}$. $E_{\mathrm{iso}}$
and $L_{\mathrm{iso}}$ are estimated in the 45-450~keV energy range.
Redshifts are provided in following references: (1) \citet{2016ApJ...829....7L};
(2) \citet{2016ApJ...817....7P};
(3) Greiner's online GRB table;
(4) the Swift GRB table: \href{http://swift.gsfc.nasa.gov/archive/grb_table.html/}{http://swift.gsfc.nasa.gov/archive/grb\_{}table.html/}.}
\end{deluxetable}

\clearpage

\begin{deluxetable}{lccc}
\tabletypesize{\scriptsize}
\tablewidth{0pt}
\tablecaption{Table of LGRB Hosts\label{tab2}}
\tablehead{
\colhead{GRB} & \colhead{$z$} & \colhead{$log(Z/\mathrm{Z_{\odot}})$} & \colhead{Ref.}}
\startdata
GRB131231A & 0.6427 & -0.24 & 1\\
GRB131103A & 0.596 & -0.21 & 1\\
GRB130925A & 0.3483 & 0.04 & 1, 3\\
GRB130702 & 0.145 & -0.5 & 3\\
GRB130603B & 0.3568 & -0.25 & 3\\
GRB130427A & 0.3401 & -0.12 & 1, 3\\
GRB120722A & 0.959 & -0.21 & 1\\
GRB120714B & 0.3985 & -0.3 & 1\\
GRB120422A & 0.2826 & -0.3 & 1\\
GRB111209A & 0.677 & -0.74 & 1, 3\\
GRB110918A & 0.9843 & 0.24 & 1, 3\\
GRB100816A & 0.8048 & 0.06 & 1\\
GRB100621A & 0.5426 & -0.17 & 1, 2\\
GRB100508A & 0.5201 & -0.01 & 1\\
GRB100418A & 0.6235 & -0.17 & 1, 3\\
GRB100206 & 0.4068 & 0.51 & 3\\
GRB091127 & 0.4903 & -0.29 & 1, 2, 3\\
GRB091018 & 0.971 & 0.09 & 1, 2\\
GRB090424 & 0.5445 & 0.19 & 2, 3\\
GRB081109 & 0.9785 & 0.06 & 1\\
GRB081007 & 0.5294 & -0.37 & 2\\
GRB080916A & 0.688 & -0.25 & 2\\
GRB080430 & 0.767 & -1.09 & 2\\
GRB071227 & 0.381 & -0.2 & 3\\
GRB071112C & 0.821 & -0.79 & 2\\
GRB070612 & 0.671 & -0.4 & 3\\
GRB061021 & 0.3453 & -0.21 & 2\\
GRB060912A & 0.9362 & -0.08 & 1, 2\\
GRB060614A & 0.125 & -0.23 & 2\\
GRB060505 & 0.0889 & -0.22 & 3\\
GRB051117B & 0.4805 & 0.31 & 1\\
GRB051022A & 0.8061 & -0.2 & 1, 3\\
GRB051016B & 0.9358 & -0.42 & 1\\
GRB050826 & 0.296 & 0.14 & 3\\
GRB050824 & 0.8277 & -0.58 & 1\\
GRB050416A & 0.6542 & -0.23 & 1, 2\\
GRB031203 & 0.1055 & -0.64 & 3\\
GRB020405 & 0.691 & -0.2 & 3\\
GRB010921 & 0.451 & -0.6 & 3\\
GRB991208 & 0.706 & -0.67 & 3\\
GRB980703 & 0.966 & -0.2 & 3\\
GRB980425 & 0.0085 & -0.5 & 3\\
\enddata
\tablerefs{(1) \citet{2015A&A...581A.125K};
(2) \citet{2016A&A...590A.129J};
(3) GHostS}
\end{deluxetable}

\end{document}